\RequirePackage{ifpdf}
\documentclass[12pt,letterpaper]{article}
\pdfoutput=1
\usepackage{jheppub}
\usepackage{epsfig}
\usepackage{enumitem}
\usepackage{bbm,amsfonts}
\usepackage{rotating,graphicx}
\usepackage{amssymb,amsmath,amsfonts,mathtools}
\usepackage{fancybox,diagbox}
\usepackage{enumerate}
\usepackage{dsfont}
\usepackage{verbatim}
\usepackage{wrapfig}
\usepackage{slashed}
\usepackage{shuffle}
\usepackage{braket}
\usepackage{pdflscape}
\usepackage{accents}
\usepackage{afterpage}

\author{Marco S. Bianchi}
 
\affiliation[a]{Facultad de Ingeniería, Universidad San Sebastián, Santiago, Chile}

\emailAdd{marco.bianchi@uss.cl}
  
\title{Tracing Transcendentality in Protected Correlators of $\mathcal{N}=4$ SYM} 

\abstract{
We study two-point functions of protected scalar operators in ${\cal N}=4$ SYM, focusing on their transcendentality properties in dimensional reduction, where quantum corrections are subleading in the regulator. We compute the correlators explicitly through two loops and operators up to classical dimension 10, for all trace structures. The one-loop correction is universal. At two loops, we find a controlled partial breaking of uniform transcendentality for higher-dimensional operators, which can be cancelled by suitable combinations of correlators in a fully predictable way. A main result is a complete planar extrapolation for two-loop correlators at arbitrary dimension and trace structure, whose dependence is entirely controlled by the number of stress-tensor multiplet factors in the operator. The perturbative results agree with localization predictions in all cases where comparisons are possible.}

\def\Tr{\textrm{Tr}}

\newcommand{\masterintegralpic}[1]{\,\raisebox{\dimexpr\fontdimen22\textfont2-.5\height\relax}{\includegraphics[scale=0.1]{pictures/NUT#1.png}}}
\newcommand{\masterintegralpicN}[1]{\,\raisebox{\dimexpr\fontdimen22\textfont2-.5\height\relax}{\includegraphics[scale=0.1]{pictures/N#1.png}}}
\newcommand{\loopintegralpic}[1]{\,\raisebox{\dimexpr\fontdimen22\textfont2-.5\height\relax}{\includegraphics[scale=0.1]{pictures/#1.png}}}
\newcommand{\masterintegralpicextra}{\,\raisebox{\dimexpr\fontdimen22\textfont2-.5\height+0.2pt\relax}{\includegraphics[scale=0.1]{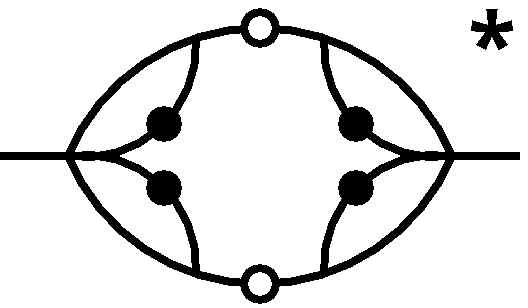}}}
\numberwithin{equation}{section}

\newlength{\dhatheight}

\begin{document}

\maketitle
\allowdisplaybreaks

\section{Introduction}

Maximally supersymmetric Yang--Mills theory provides a remarkably constrained laboratory for perturbative and non-perturbative quantum field theory. Its high degree of symmetry underlies integrability in the planar limit \cite{Beisert:2010jr}, powerful non-renormalization theorems for protected operators \cite{Lee:1998bxa,Intriligator:1999ff,Eden:1999gh,Arutyunov:2001qw,Heslop:2001gp}, and an exceptionally rich structure of correlation functions \cite{Eden:1998hh,Eden:1999kh,Eden:2000mv,Dolan:2001tt,Dolan:2004iy}. These features make protected correlators a natural arena in which to compare direct Feynman-integral calculations, symmetry-based expectations, and localization-inspired results.

A recurring theme in the study of scattering amplitudes and correlation functions in $\mathcal{N}=4$ SYM is the appearance of functions and constants of uniform transcendental weight. This structure has been observed extensively in amplitudes \cite{Bern:2005iz,Goncharov:2010jf}, Wilson loops \cite{DelDuca:2009au,DelDuca:2010zg}, and form factors \cite{vanNeerven:1985ja,Bork:2010wf,Gehrmann:2011xn,Brandhuber:2012vm,Brandhuber:2014ica,Banerjee:2016kri,Huber:2019fxe,Lin:2020dyj,Agarwal:2021zft,Lee:2021lkc}. It is often viewed as part of the hidden simplicity of the theory, closely related to integrability \cite{Beisert:2010jr,Arkani-Hamed:2012zlh,Kotikov:2003fb,Kotikov:2004er,Eden:2006rx,Beisert:2006ez,Marboe:2014sya,Marboe:2016igj,Kniehl:2021ysp} and to the use of bootstrap strategies for amplitudes and form factors \cite{Dixon:2011pw,Dixon:2011nj,Dixon:2013eka,Drummond:2014ffa,Caron-Huot:2016owq,Dixon:2016nkn,Drummond:2018caf,Caron-Huot:2019vjl,Dixon:2020cnr,Dixon:2020bbt,Dixon:2022rse,Dixon:2022xqh}. From the Feynman-integral perspective, this property is closely related to the existence of special bases of master integrals, often chosen so that differential equations or integral expansions take a canonical form \cite{Henn:2013pwa}. Uniform transcendentality is not guaranteed for generic observables,  nevertheless it has repeatedly served as an organizing principle and as a useful diagnostic of hidden simplicity in perturbative results.

In \cite{Bianchi:2023llc}, the two-point function of protected dimension-2 operators was found to exhibit striking transcendentality properties in dimensional regularization (and an analogous pattern emerges in ABJM theory \cite{Aharony:2008ug,Aharony:2008gk} for similar operators \cite{Bianchi:2024nah}). More recently, a uniformly transcendental basis for the relevant four-loop propagator-type master integrals was constructed in \cite{Bianchi:2025sjc}. These developments motivate a broader question: to what extent do similar structures persist for protected operators of higher classical dimension, where multi-trace operators, richer color factors, and more complicated mixing patterns are unavoidable?

The purpose of this paper is to address this question by computing two-point functions of protected scalar operators in $\mathcal{N}=4$ SYM through two loops for operators of arbitrary trace structure, with explicit data generated up to dimension 10. We work in dimensional reduction, which is the natural supersymmetry-preserving version of dimensional regularization for the present purposes \cite{Siegel:1979wq,Capper:1980qk,Velizhanin:2008rw,Chakraborty:2026hdv}. The calculation is performed in momentum space, reduced to propagator master integrals by IBP identities and Laporta reduction \cite{Chetyrkin:1981qh,Tkachov:1981wb,Laporta:1996mq,Laporta:2001dd}, and then transformed to position space, where the two-point functions are most naturally expressed. The full results are available in electronic format as \texttt{Mathematica} files attached to this submission.

The resulting data also provide perturbative input for comparison with localization-based predictions \cite{AleCongkao}. In agreement with non-renormalization theorems for protected operators \cite{Lee:1998bxa,Intriligator:1999ff,Eden:1999gh,Arutyunov:2001qw,Heslop:2001gp}, the quantum corrections are subleading in the dimensional regulator $\epsilon$. This makes contact with localization on spheres of generic dimension \cite{Minahan:2015jta,Minahan:2017jkg,Gorantis:2017vra,Minahan:2015cva}, where the dimensional continuation provides a natural framework for comparison.

Our main findings are as follows. At one loop, the correction to the two-point function of any protected operator factorizes in a universal way, with a simple dependence on the classical dimension, as shown in \eqref{eq:oneloopratio}--\eqref{eq:oneloopf}. At two loops, the dimension-3 single-trace correlator remains uniformly transcendental after a natural normalization, see \eqref{eq:f3}. Starting at dimension 4, individual correlators generally cease to be uniformly transcendental, but the breaking is highly constrained. In particular, suitable combinations of correlators, involving single- and multi-trace operators, restore uniform transcendentality; the corresponding subtraction pattern is encoded in \eqref{eq:v} and the resulting combination is given in \eqref{eq:Unmaster}--\eqref{eq:Unexpansion}. Beyond these cancellations, we obtain a compact all-$n$ planar conjecture for diagonal two-loop correlators of arbitrary trace structure, summarized in \eqref{eq:fullplanar}. If the trace structure is represented by a partition $\Lambda$, the full master-integral expression and its $\epsilon$ expansion depend on $\Lambda$ only through the counting function $m_2(\Lambda)$, defined in \eqref{eq:number-of-twos}, namely the number of factors of $\Tr(X^2)$. This provides a simple organizing principle for the planar trace-structure dependence, in line with the tower organization proposed in \cite{Brown:2023zbr}.

The paper is organized as follows. The first two sections fix conventions, define the protected operators and trace basis, and derive the universal one-loop result. In section \ref{sec:two-loop} we describe the two-loop setup and the effective-vertex organization of the calculation. We then discuss the dimension-3 and dimension-4 examples in detail, since they display respectively the persistence and the first breaking of uniform transcendentality. The following section presents extrapolations to general operator dimension and collects compact formulae for special trace structures. We then describe uniformly transcendental subtractions and the combinatorial pattern controlling them. Finally, section \ref{sec:full-planar} gives the full planar extrapolation for diagonal two-loop correlators, exhibiting the dependence on the trace structure through $m_2(\Lambda)$ alone, before we conclude with open directions.

\section{Definitions}
We work in ${\cal N}=4$ SYM with gauge group $SU(N)$, dimensionally regularized in $d=4-2\epsilon$.
Dimensional reduction \cite{Siegel:1979wq,Capper:1980qk} is adopted throughout the paper. At the perturbative order considered here, this scheme preserves supersymmetry, as recently established on firmer grounds in \cite{Chakraborty:2026hdv} and previously argued in \cite{Velizhanin:2008rw}.
To address natural questions about the choice of regulator and possible alternatives, we summarize the reasons for using dimensional reduction.
The two-point functions considered in this work are finite and, in fact, of order $\epsilon$, as implied by non-renormalization theorems \cite{Intriligator:1999ff}. In perturbation theory, this desirable property holds provided the regularization scheme respects the hypotheses of such theorems. Dimensional reduction is therefore a suitable choice for exposing this structure.
Another goal of this work is to provide data for comparison with localization-based calculations \cite{AleCongkao}. Since localization also relies on supersymmetry, a supersymmetry-preserving regularization scheme is required for a meaningful comparison.
Finally, we aim to assess the transcendentality properties of these two-point functions, motivated by previous results that revealed hints of uniform transcendentality for the two-point function of dimension-2 protected operators \cite{Bianchi:2023llc}. Those results were obtained using dimensional reduction, so it is natural to continue working in this scheme. The two-loop calculations presented below are also technically feasible within dimensional regularization, although it might be possible to perform them in other regularization schemes as well.

We now introduce the protected scalar operators whose two-point functions will be studied. For a fixed classical dimension $n$, the operators are built out of $n$ adjoint scalar fields, with gauge indices contracted according to a chosen color-trace structure. We denote a generic element of this trace basis by
\begin{equation}\label{eq:operators}
    O_{n,i} = \underbrace{X^{a_1} \dots X^{a_n}}_{n} \, \times \,\text{trace basis}_i \left( T^{a_1} \dots T^{a_n} \right)
\end{equation}
where the label $i$ distinguishes the different admissible trace structures. For dimensions 2 and 3, only single-trace operators are possible. Starting at dimension 4, multi-trace operators also appear. We classify the trace-basis elements by partitions of $n$ whose parts have minimum size 2, and order them first by the number of traces and then lexicographically. For instance,
\begin{equation}
    O_{10,10} = \Tr(X X X X) \Tr^3(X X) \quad,\quad  O_{10,11} = \Tr^2(X X X) \Tr^2(X X)  \quad \text{etc.}
\end{equation}
The complete list of basis elements is stored in the attached file \texttt{data.m} and can be retrieved by calling \texttt{operator[n,i]}.

Let us stress that, in this work, the term protected refers to the chiral sector from which the operators are constructed, rather than to a basis of orthogonal superconformal primary operators. The trace structures in \eqref{eq:operators} provide a convenient holomorphic multi-trace basis for perturbation theory. At finite $N$, operators with the same quantum numbers may mix, and the trace basis need not coincide with the basis of normalized superconformal primaries. Since our main goal is to study the perturbative two-point functions and their transcendentality properties in this explicit trace basis, we do not perform an orthogonalization or diagonalization of the correlator matrix.

The diagrams are more conveniently evaluated perturbatively in momentum space, where the integrals are amenable to automated IBP reductions to master integrals \cite{Chetyrkin:1981qh,Tkachov:1981wb,Laporta:1996mq,Laporta:2001dd} and to subsequent expansions in $\epsilon$.
For both tasks, we used \texttt{Forcer} \cite{Ruijl:2017cxj} in \texttt{Form} \cite{Vermaseren:2000nd,Ruijl:2017dtg,Davies:2026cci}. We then Fourier transform the results to position space, which provides the natural variables for two-point functions. More technical details on the computation are given in section \ref{sec:two-loop}.

To fix our notation, we present the results in $x$-space.
At tree level, we write the correlator as the product of a kinematic factor, given by the product of $n$ free propagators, a combinatorial factor $n$, corresponding to the number of inequivalent Wick contractions for single-trace operators, and a suitably normalized color factor,
\begin{equation}\label{eq:tree}
\left\langle O_{n,i}(x) O_{n,j}(0) \right\rangle^{(0)} = 
 \Pi^{n}(x,\epsilon)\, 
n
\times \text{color factor}_{ij}
\end{equation}
where
\begin{equation}
\Pi(x,\epsilon) \equiv \frac{\Gamma\left(1-\epsilon\right)}{(4\pi^2)\, x^2} \left(\pi\, x^2\right)^{\epsilon}
\end{equation}
For single-trace operators in the large-$N$ limit, this becomes
\begin{equation}\label{eq:tree11}
\left\langle O_{n,1}(x) O_{n,1}(0) \right\rangle^{(0)} = 
 \Pi^{n}(x,\epsilon)\, 
n\, N^n + O\left(N^{n-2}\right)
\end{equation}
For subleading corrections and other trace structures, we refer to the attached files for the full list of color factors for operators up to $n=10$, for which we have computed the corresponding quantum corrections up to second order.

For the loop corrections presented in the following sections, it is convenient to factor out the tree-level result and define the perturbative expansion as
\begin{equation}
   G_{n,i,j} \equiv \left\langle O_{n,i}(x) O_{n,j}(0) \right\rangle = G_{n,i,j}^{(0)}\left(1+\lambda F_{n,i,j}^{(1)}+\lambda^2 F_{n,i,j}^{(2)}+\mathcal{O}\left(\lambda^3\right)\right)
\end{equation}
where
\begin{equation}
   \lambda \equiv \frac{g^2 N}{16\pi^2}\, \left(e^{\epsilon \gamma_E} \pi\, x^2\right)^\epsilon
\end{equation}
which systematically absorbs powers of $\pi^\epsilon$, $\left(x^2\right)^{\epsilon}$, and $e^{\gamma_E}$, thereby simplifying the expressions for $F_{n,i,j}$.

\section{One loop}
At the level of Feynman integrals, this contribution is obtained by inserting the one-loop correction to a dimension-2 two-point function into a larger diagram.
No other contributions are possible. The color structure of these diagrams is also elementary, since the one-loop correction is always proportional to the tree-level one by a factor of $N$, as can be proven by directly inspecting the relevant Feynman-diagram algebra. 
In the simplest topology, one-loop corrections to the scalar propagators naturally give the same color structure as at tree level, since these self-energies do not alter the color indices of the operators.
The next class of diagrams involves a gluon exchange. Working out the color algebra produces two different color traces with equal coefficients. This implies that, after summing over all scalar contractions with the operators, the color factor is the same as in the tree-level calculation.
This result is also enforced by gauge invariance, since this diagram contains a $\xi$-dependent term in general $\xi$ gauge, which must be exactly cancelled by the corresponding term from the gluon-exchange part of the self-energy diagrams. Hence their color coefficients must be equal, and the latter is trivially identical to the tree-level structure. The final class is given by a four-scalar vertex. In this case as well, explicitly working out the color algebra leads to the same combination of traces, with the same coefficients as in the gluon-exchange case.

The problem then boils down to the combinatorics of Wick contractions.
A generic expression for the final correlation function ratio can be written in terms of the following integrals:
\begin{align}\label{eq:oneloopint}
   F_n^{(1)} &= -\frac{\left(x^2\right)^{-\epsilon}}{\Pi^2(x,\epsilon)}\, \frac{e^{-3\gamma_E\epsilon}}{4^{\frac32 d -4}}\,\int \frac{d^d p}{\pi^{d/2}}\, e^{i p \cdot x}\, 
   \frac{n}{\epsilon ^2}\, \bigg(
   2 (2-3 \epsilon) (1-3 \epsilon) \loopintegralpic{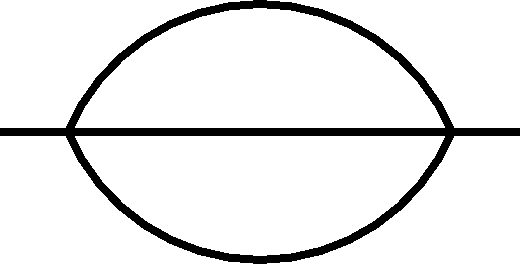}\, \frac{1}{p^2}
   \nonumber\\& \qquad\qquad\qquad\qquad\qquad\qquad
   +(1-2\epsilon) \epsilon  \loopintegralpic{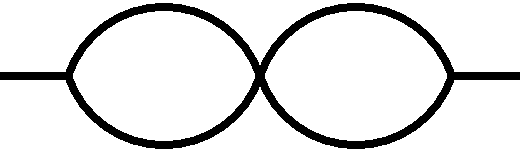}  
   \bigg)
\end{align}
which can be expressed in terms of gamma functions. Their $\epsilon$-dependent factors arise from the IBP reduction.
The final position-space correlator is obtained by Fourier transforming the integrals and adding $(n-2)$ free propagators. The power of 4 in front of the integral cancels the corresponding factor generated by the Fourier transform, simplifying the formulae below. Dividing by the tree-level expression gives \eqref{eq:oneloopint}. The power of $e^{\gamma_E\epsilon}$ eliminates spurious factors in the following results.

Analyzing the transcendentality properties of the integrals in this expression, whose evaluation is recalled in appendix \ref{app:expansions}, one finds that multiplying them by $\frac{1-2 \epsilon}{1-3 \epsilon}$ enforces uniform transcendentality. 
Therefore, in general we can write
\begin{equation}
F_n^{(1)} = 
-\frac{1-3 \epsilon}{1-2 \epsilon} \, \, 
n\,
f^{(1)}(\epsilon)
\end{equation}
The function $f^{(1)}$ is defined as
\begin{equation}\label{eq:oneloopratio}
   f^{(1)}(\epsilon) = e^{-\gamma_E\epsilon}\, \frac{2 \Gamma (1-\epsilon ) \left(\frac{\cos (\pi  \epsilon ) \Gamma (1-3 \epsilon) \Gamma (1+\epsilon)}{\Gamma (1-2 \epsilon )}-1\right)}{\epsilon ^2}
\end{equation}
and is manifestly uniformly transcendental, with a transcendental weight 2 expansion
\begin{align}\label{eq:oneloopf}
    f^{(1)}(\epsilon) &= 12 \zeta_3 \epsilon +18 \zeta_4 \epsilon ^2+(6 \zeta_2 \zeta_3+84 \zeta_5) \epsilon
   ^3+\left(40 \zeta_3^2+\frac{783 \zeta_6}{4}\right) \epsilon^4\nonumber\\&+\left(\frac{483 \zeta_3
   \zeta_4}{4}+42 \zeta_2 \zeta_5+588 \zeta_7\right) \epsilon ^5+O\left(\epsilon^6\right)
\end{align}
In conclusion, \eqref{eq:oneloopratio}--\eqref{eq:oneloopf} give the full one-loop correction in $d=4-2\epsilon$ to any two-point function of protected operators of the form \eqref{eq:operators}, for generic classical dimension $n$. In particular, the dependence on $n$ is elementary and is entirely captured by the overall factor of $n$ in the expression above.

As a first important result, we stress that this expression is in complete agreement with the first few orders in the $\epsilon$ expansion predicted by localization \cite{AleCongkao}.

\section{Two loops}\label{sec:two-loop}

\begin{figure}[htbp]
   \centering
   \[
   \vcenter{\hbox{\includegraphics[width=0.14\textwidth]{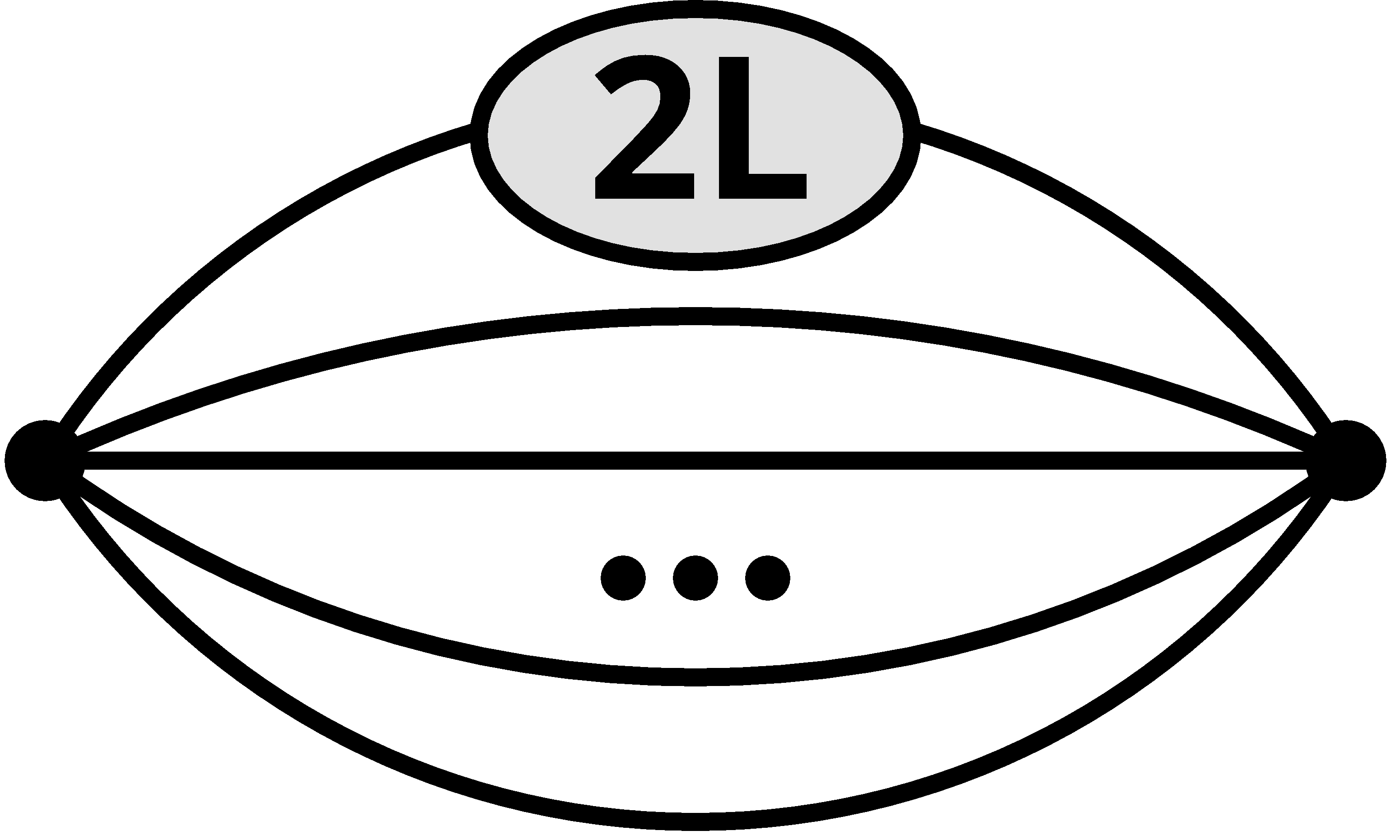}}}
   \,\,+\,\,\vcenter{\hbox{\includegraphics[width=0.14\textwidth]{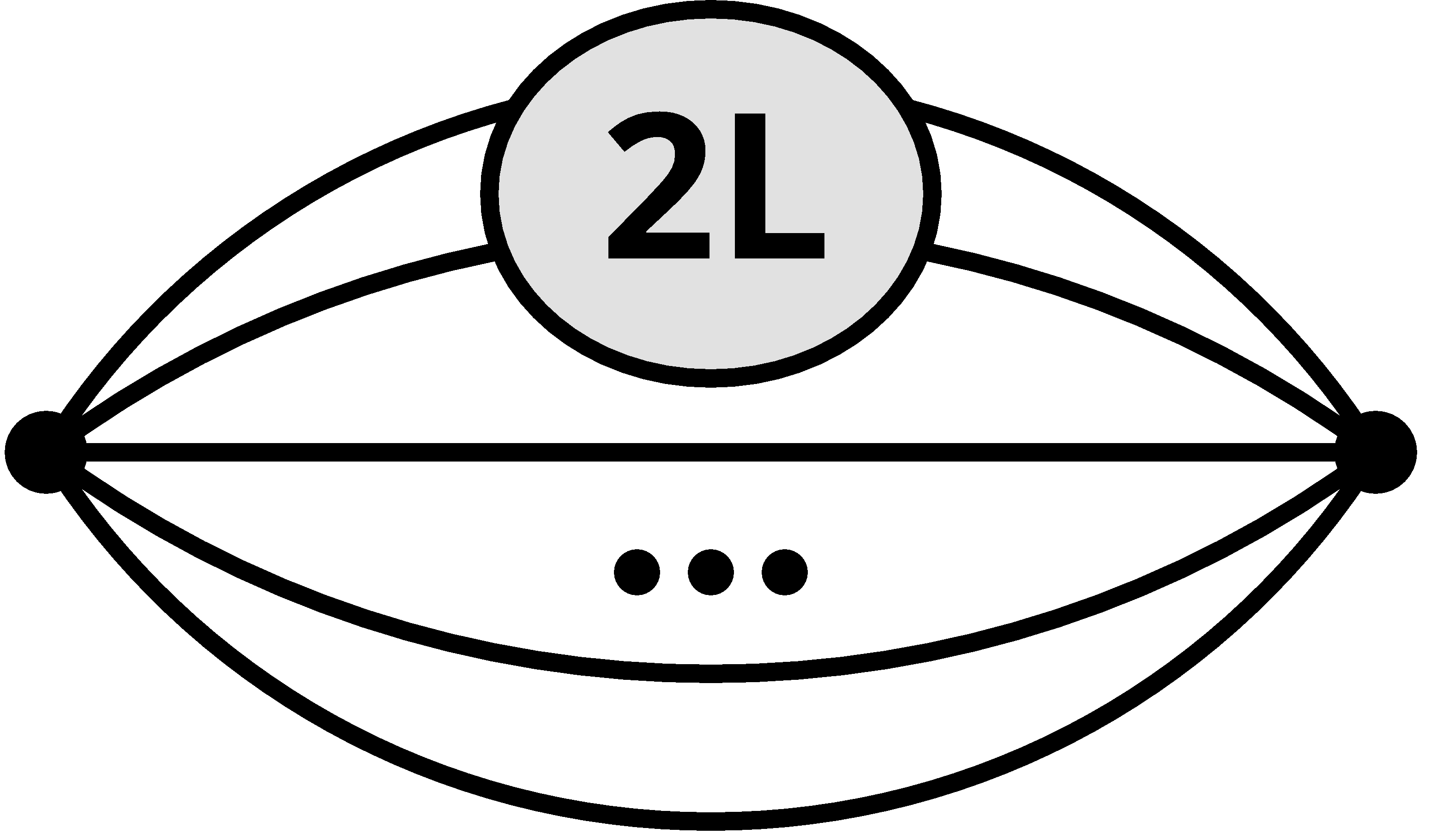}}}
   \,\,+\,\,\vcenter{\hbox{\includegraphics[width=0.14\textwidth]{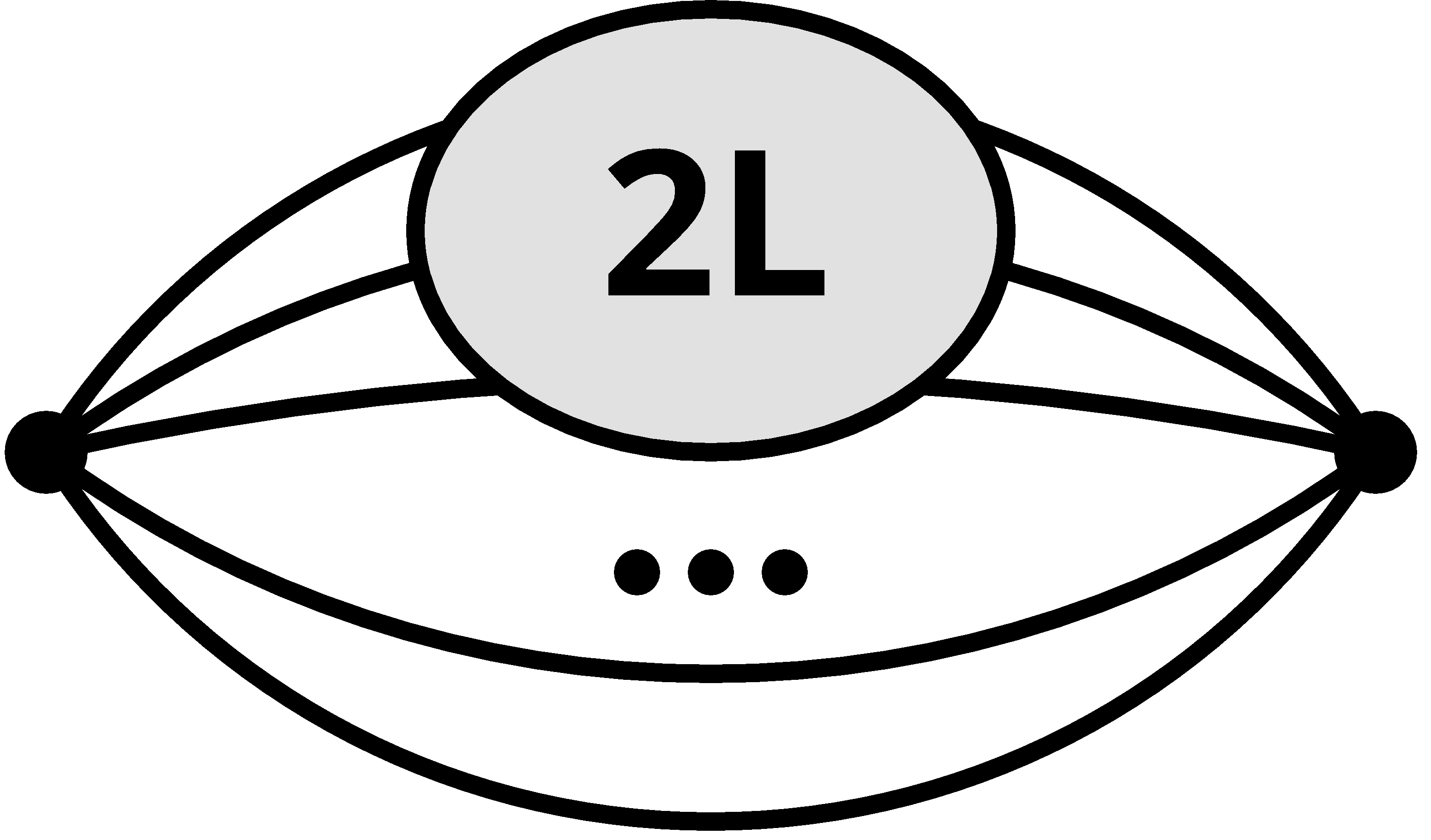}}}
   \,\,+\,\,\vcenter{\hbox{\includegraphics[width=0.14\textwidth]{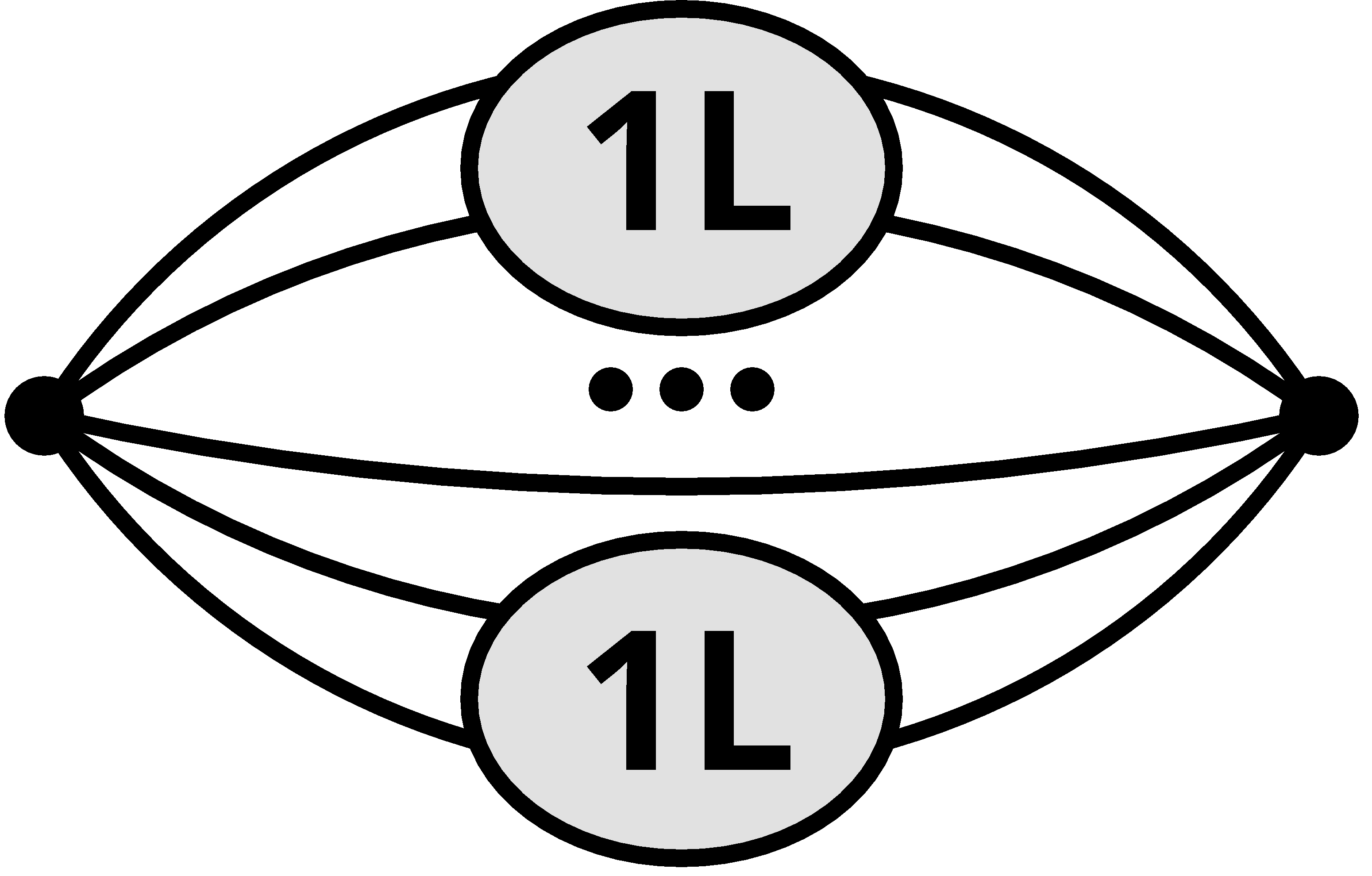}}}
   \,\,=\,\,\vcenter{\hbox{\includegraphics[width=0.14\textwidth]{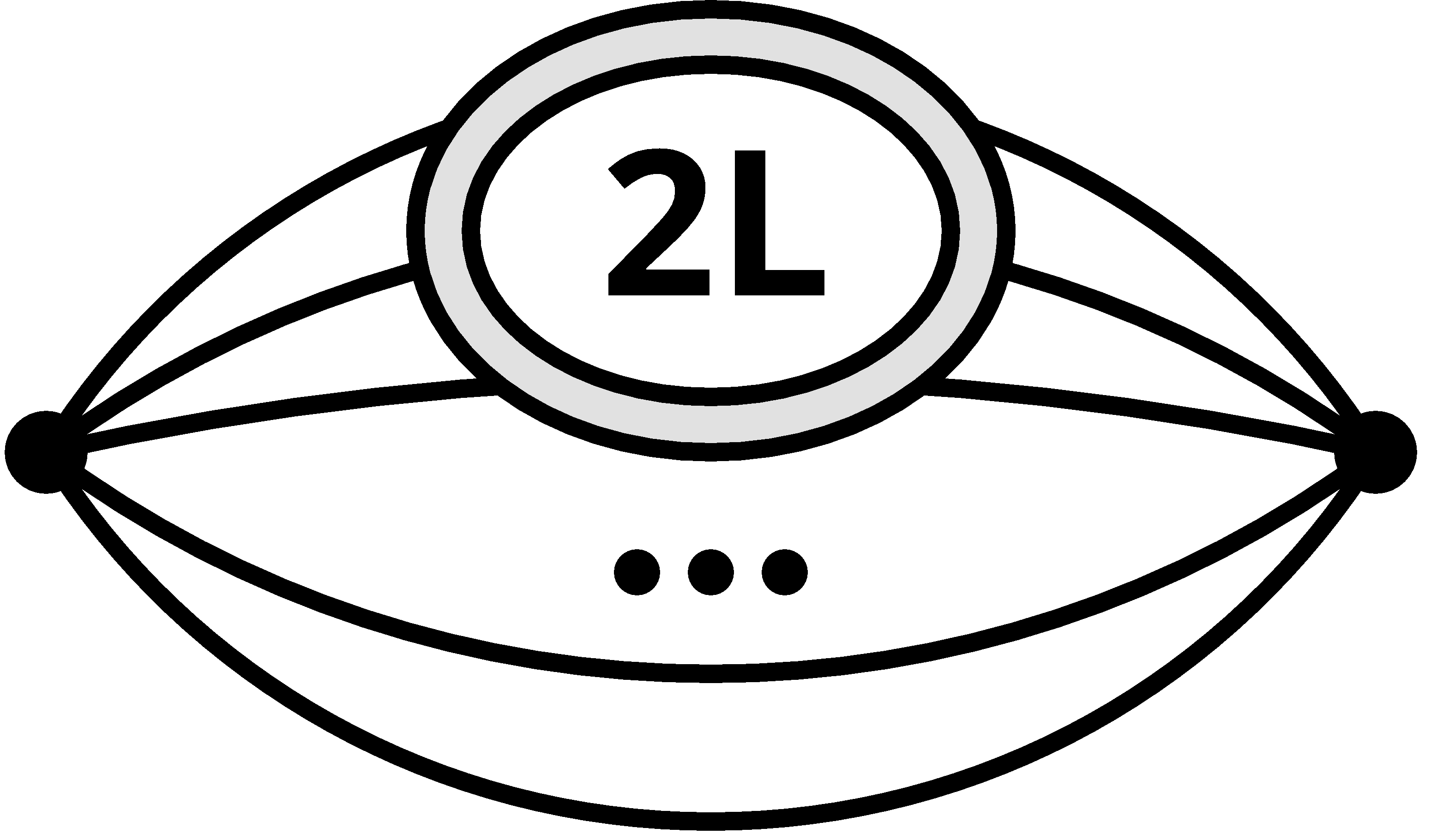}}}
   \]
   \caption{Schematic construction of the effective vertex used in the two-loop calculation.}
   \label{fig:effective}
\end{figure}

The two-loop calculation entails substantially more computational effort.
To reach higher operator dimensions, we optimized the calculation as follows.
There are about 100 diagram topologies contributing to the calculation, which we generated with \texttt{QGRAF} \cite{Nogueira:1991ex}. These can be grouped according to the number of external operator legs they involve.
The simplest ones are essentially self-energy corrections, represented by the first term in figure \ref{fig:effective}. They are followed by contributions involving only two external legs, which are the diagrams appearing in two-point functions of dimension-2 operators and are represented by the second term in figure \ref{fig:effective}. For dimension-3 operators, new two-loop diagrams involving three external operator legs are present, represented by the third term in figure \ref{fig:effective}.
In momentum space, these diagrams already give rise to four-loop integrals at second order in perturbation theory. This marks a significant increase in complexity compared with dimension-2 operators, for which two-point functions at perturbative order $l$ involve at most $(l+1)$-loop integrals. This is the main reason why we were not able to push the calculation to third order for higher-dimensional operators. For instance, for dimension-3 operators at third order, genuine six-loop momentum integrals would be needed, which are beyond our present computational capabilities.

Finally, starting with dimension-4 operators, new diagrams appear which would in principle correspond to five-loop momentum integrals, but factorize into products of two separate two-loop contributions and are therefore easy to evaluate. This factorized contribution is represented schematically by the fourth term in figure \ref{fig:effective}.

For operators of higher dimension $n$, the momentum-space contributions would in principle look like $(n+1)$-loop integrals, but their non-trivial part is always one of the cases mentioned above, multiplied by additional free propagators.

Since we performed the IBP reduction with \texttt{Forcer}, whose capability is limited to four-loop momentum integrals, we adopted the following strategy.
First, we pre-evaluate an effective correction by summing all non-trivial contributions in terms of master integrals. The sum of these contributions defines the effective vertex shown on the right-hand side of figure \ref{fig:effective}.

After this reduction, the only remaining ingredient is the color algebra, which we also reduce to a minimal set of ten inequivalent internal index contractions.
Second, we attach this effective blob to the external legs of the operators in the two-point functions.

This latter task is trivial in principle, but the combinatorics scale rapidly with the operator dimension. Moreover, the number of possible trace structures increases with the dimension, adding further computational burden.

The calculation of the color factors associated with thousands of contractions of color generators is algorithmic and straightforward in principle, but it becomes time-consuming for higher-dimensional operators because the combinatorics grow rapidly. We mitigated this by restricting the effective vertices to a small number of color contractions and by implementing symmetries such as cyclicity of traces and exchange properties when several traces of the same length are multiplied. For the remaining inequivalent contractions, the color factors were computed using the $SU(N)$ routine of the \texttt{color} package \cite{vanRitbergen:1998pn} in \texttt{Form}.

In practice, this procedure significantly sped up the calculation compared with a direct evaluation of all diagrams, which we nevertheless performed for cross-checks in lower-dimensional cases. We were able to compute the full trace basis of operators up to dimension 10 in a reasonable time, namely a few days on a 32-core workstation. We also computed a few additional trace structures at dimensions 11 and 12 to test predictions obtained by extrapolating lower-dimensional results to all $n$.

Unlike in the one-loop case, at two loops the color factors are no longer proportional to the tree-level results and contain non-trivial new information.
Two exceptions are the simplest dimension-2 case, already presented up to three loops in \cite{Bianchi:2023llc}, and the dimension-3 case, whose result we present here.

\paragraph{Making master-integral expansions uniform}

In the following, we provide expressions both in terms of master integrals and as expansions in $\epsilon$.
Master-integral expressions are more general, and by selecting a basis of uniformly transcendental master integrals we can make this property more manifest or, when it is broken, inspect its origin more closely.

The integrals are considered in momentum space, where they are $(n+l-1)$-loop momentum integrals.
This is somewhat inconvenient when comparing correlators of operators with different dimensions $n$.
However, at low loop order it is possible to separate each integral into a non-trivial part multiplied by free propagators in $x$-space.

Since we use \texttt{Forcer} for the IBP reductions, we restrict the loop integrals to four loops.
However, there are three diagram topologies that look like five-loop integrals. These are the factorized topologies represented schematically by the fourth contribution in figure \ref{fig:effective}. In each case, the main building blocks are two separate two-loop contributions of the following forms:
\begin{figure}[ht]
   \centering
   \includegraphics[width=0.15\textwidth]{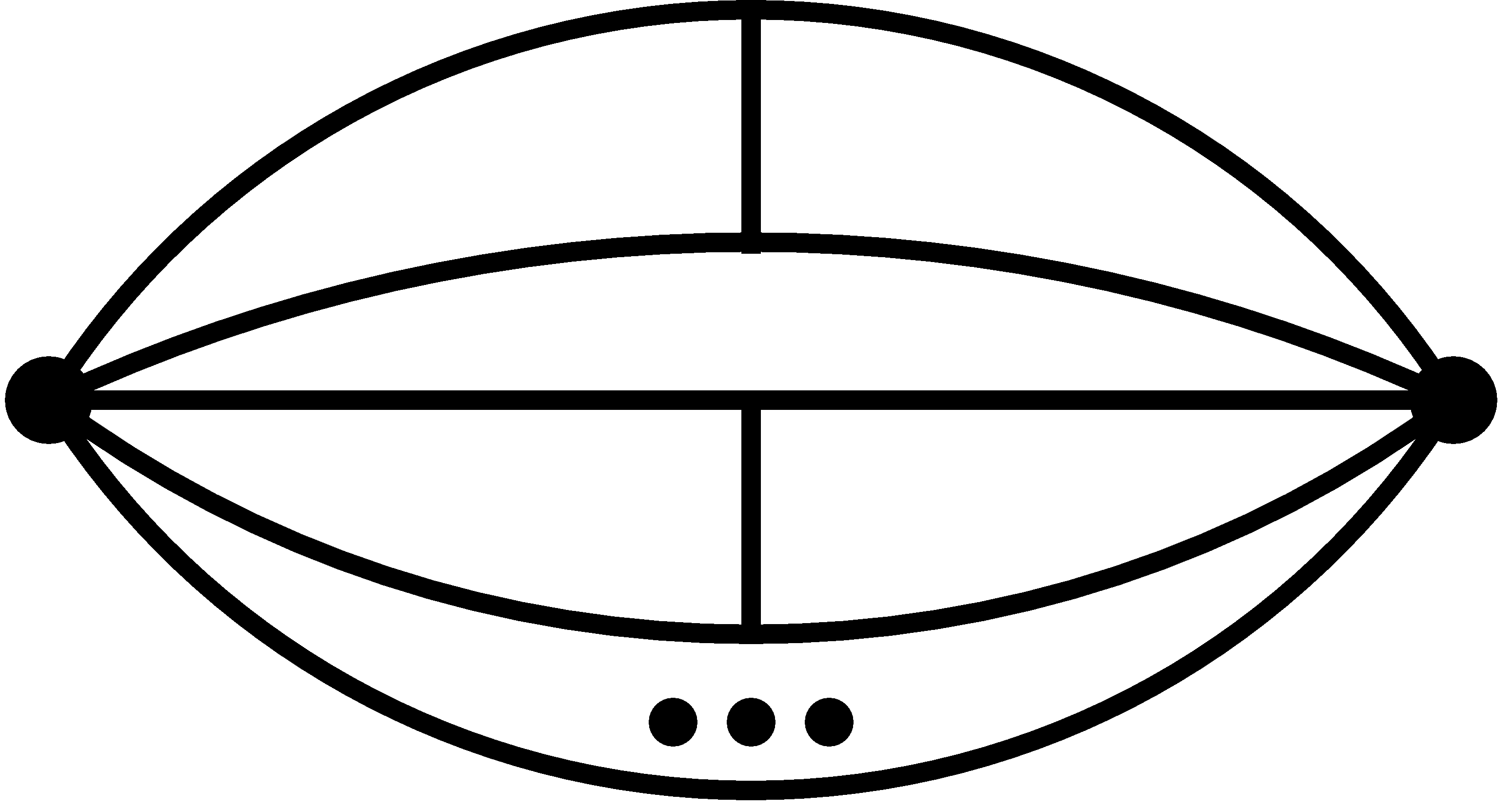}
   \hspace{0.03\textwidth}
   \includegraphics[width=0.15\textwidth]{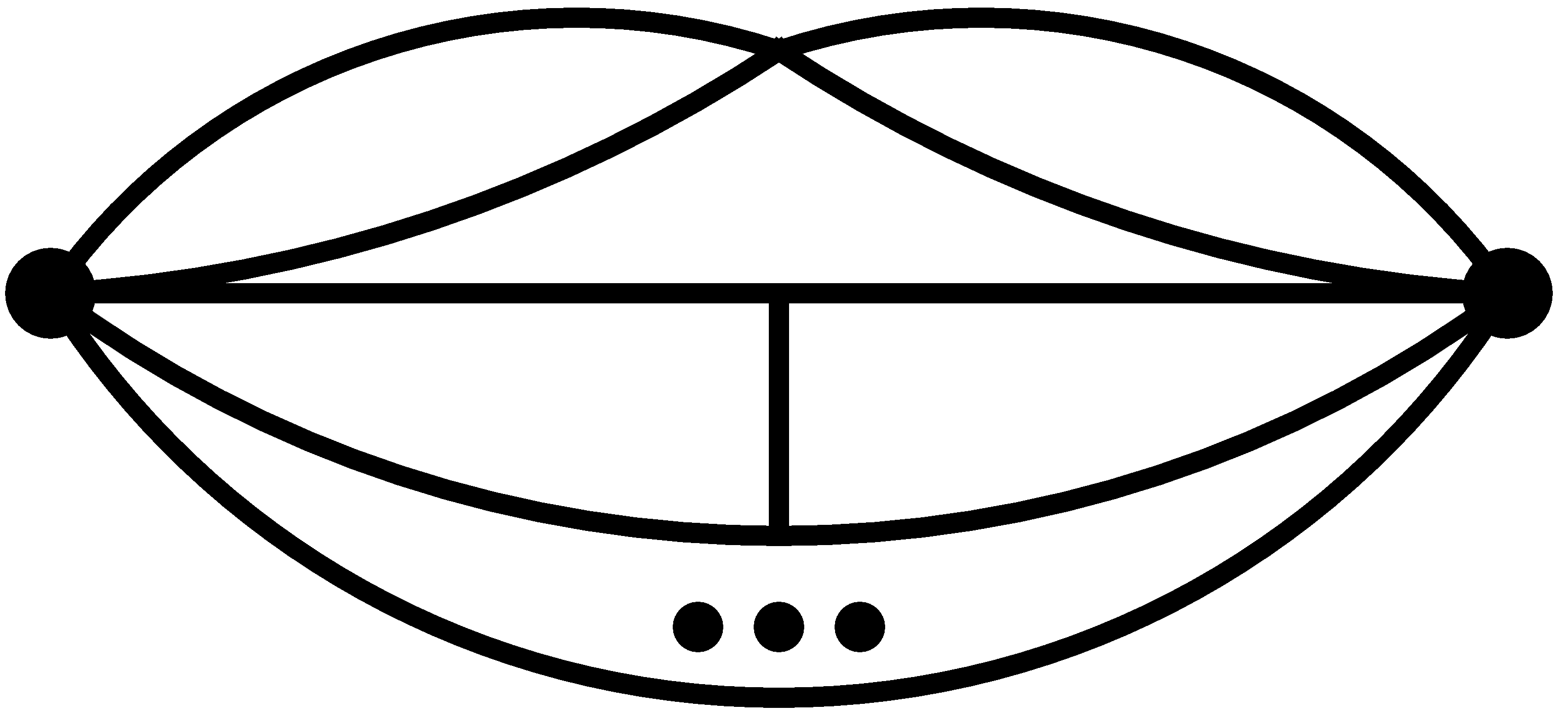}
   \hspace{0.03\textwidth}
   \includegraphics[width=0.15\textwidth]{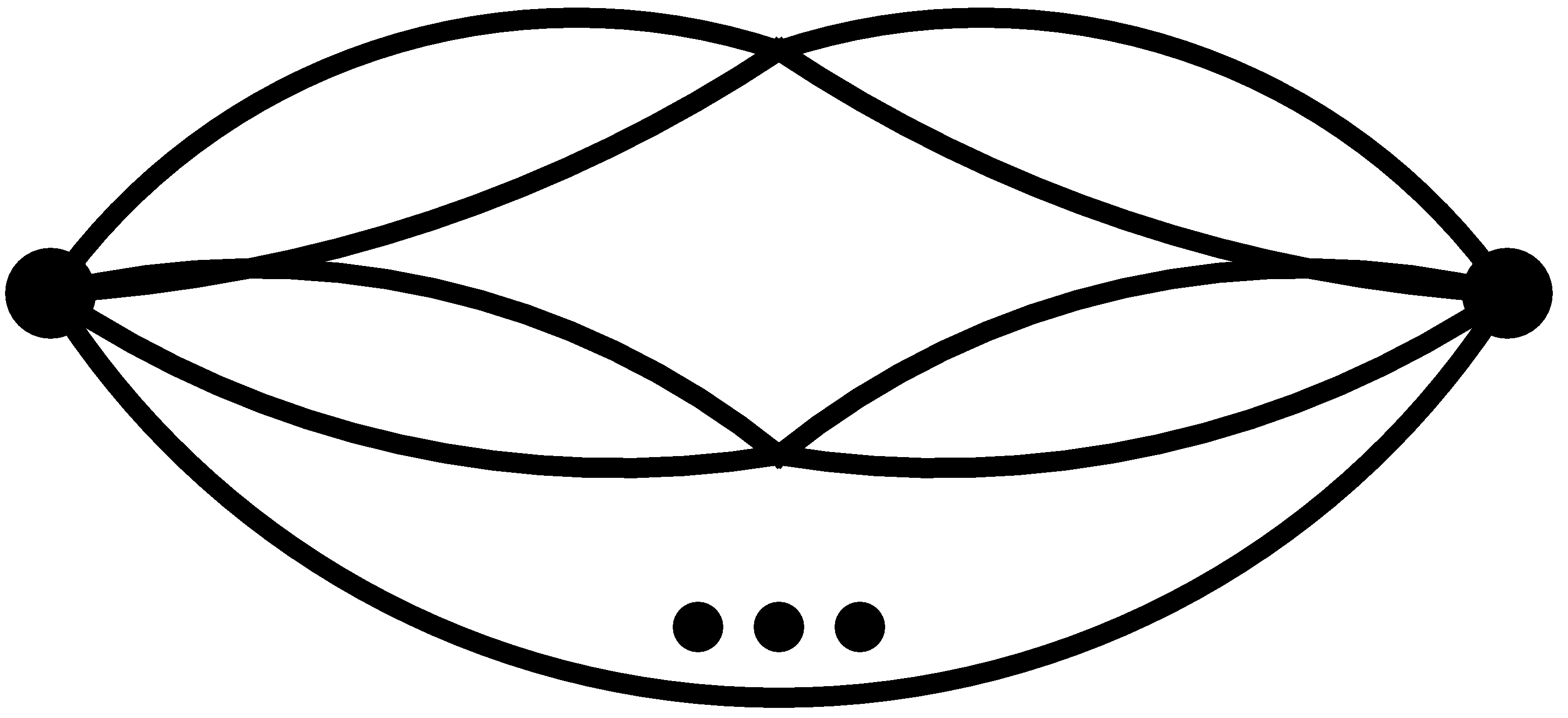}
\end{figure}

After integration by parts on the separate two-loop blocks, the relevant topologies simplify to:
\begin{figure}[ht]
   \centering
   \includegraphics[width=0.15\textwidth]{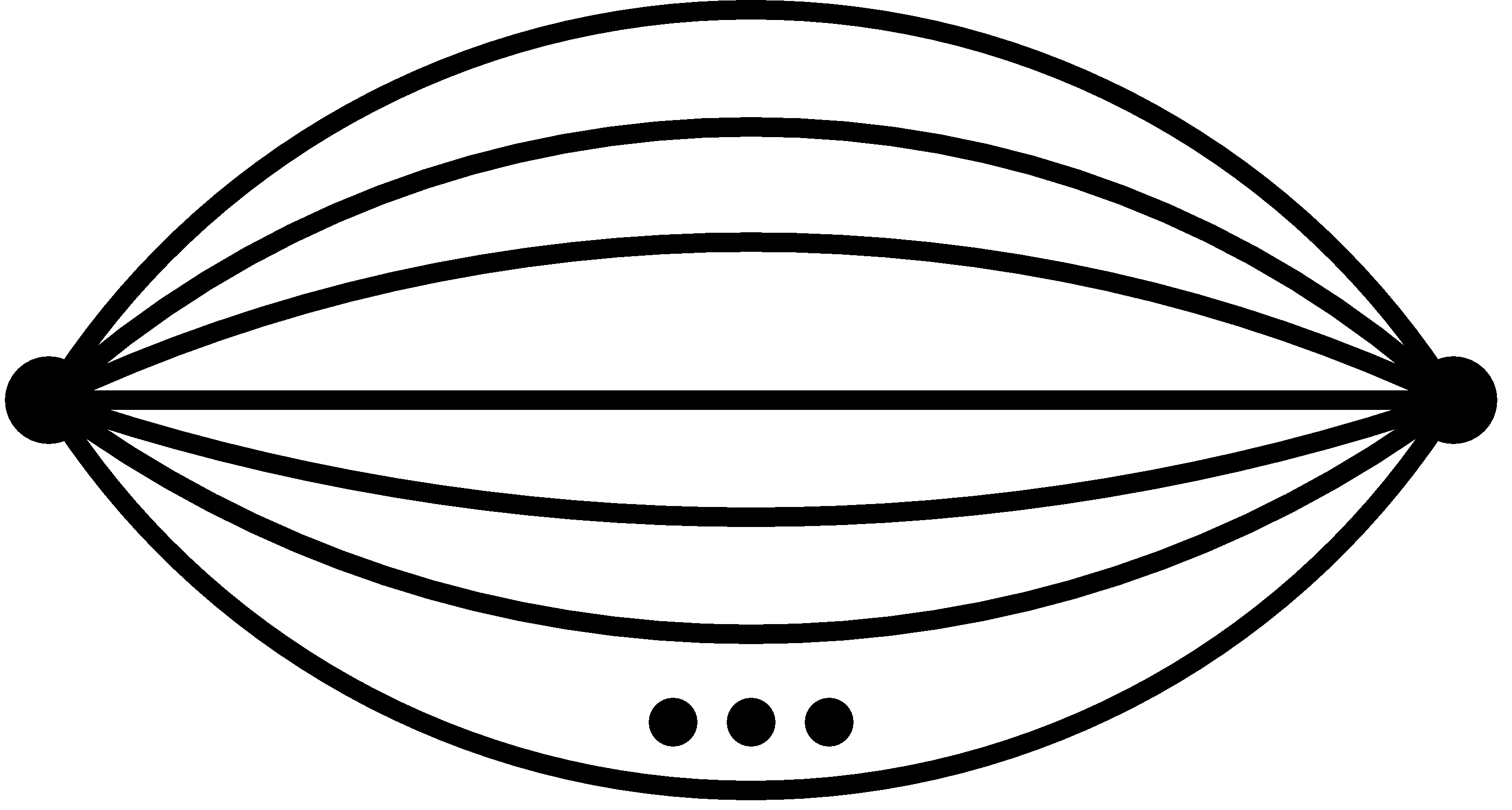}
   \hspace{0.03\textwidth}
   \includegraphics[width=0.15\textwidth]{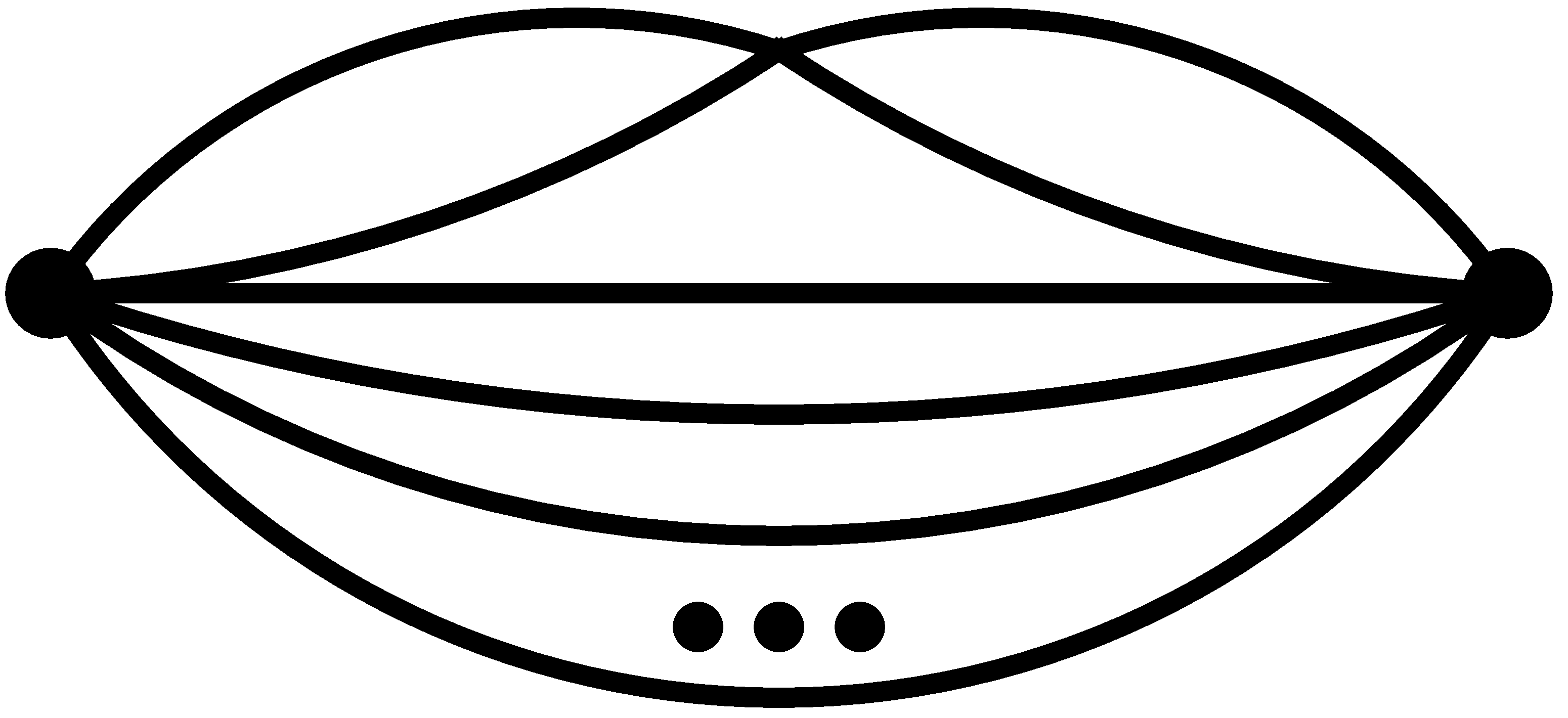}
   \hspace{0.03\textwidth}
   \includegraphics[width=0.15\textwidth]{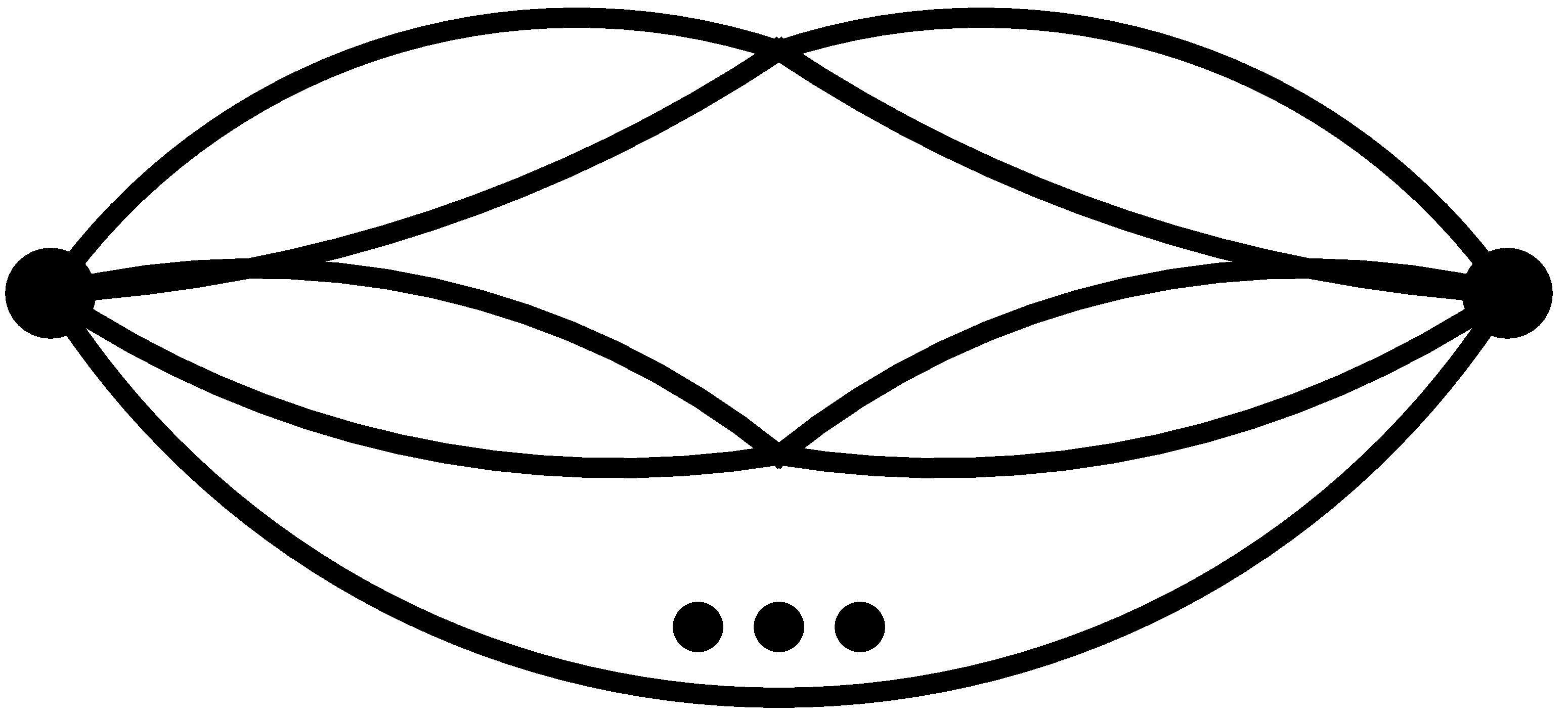}
\end{figure}

For the first two, the presence of at least one free propagator means that they can easily be reduced to the product of a four-loop integral and a free propagator, and can therefore be assimilated into the rest of the calculation.

However, the third topology is inherently five-loop, although it is easy to evaluate in terms of bubble integrals.
A uniformly transcendental version of this integral is
\begin{equation}\label{eq:extra}
   \frac{1-6\epsilon}{\epsilon} \, \raisebox{\dimexpr\fontdimen22\textfont2-.5\height+0.2pt\relax}{\includegraphics[scale=0.1]{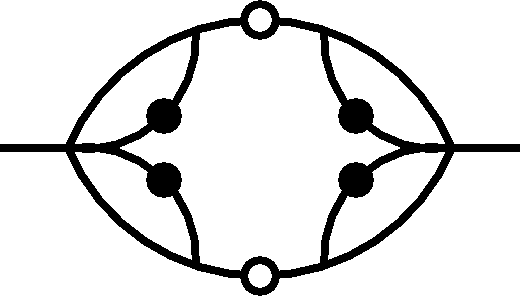}}
\end{equation}
Here and in the following integral diagrams appearing in formulae, solid dots denote squared propagators, i.e.~$(k^2)^{-2}$ factors, whereas empty dots denote numerator factors, i.e.~$k^2$, if the corresponding momentum is $k$.
The factor $(1-6\epsilon)$ enforces uniform transcendentality, while $\epsilon^{-1}$ makes the transcendental weight the same as the other basis elements.

When Fourier transforming to position space, this integral is multiplied by a factor different from the others. This is harmless, but it can obscure its combination with the other basis elements, which are organized as dimension-3 two-point-function integrals.

To keep the master-integral formulae more homogeneous, we define the following integral:
\begin{align}
   \masterintegralpicextra &\equiv \frac{2 (1-4 \epsilon)\, e^{-\gamma_E\epsilon}}{3 (1-2 \epsilon) (1-3 \epsilon) (1-6 \epsilon)}\, \int  \frac{d^d x}{(4\pi)^{d/2}}\, \frac{e^{-ipx}}{\Pi(x,\epsilon)} 
   \nonumber\\&\qquad\qquad\qquad
   \int \frac{d^d p}{\pi^{d/2}}\, e^{ip'x}\, \frac{1-6\epsilon}{\epsilon} \, \raisebox{\dimexpr\fontdimen22\textfont2-.5\height+0.2pt\relax}{\includegraphics[scale=0.1]{pictures/M5.png}}\,\bigg|_{p'}
\end{align}
We Fourier transform the integral \eqref{eq:extra}, evaluated at $p'$, to position space, remove one free-propagator factor to cast it as a dimension-3 two-point-function integral, and finally transform it back to momentum space. In this way, the extra integral is put on the same footing as the other four-loop uniformly transcendental integrals in momentum space.
The additional rational factors of $\epsilon$ in front restore uniform transcendentality, which is partially obscured by the Fourier transforms.

This complication is designed to make the perturbative results below more compact.

With the conventions of appendix \ref{app:expansions}, the Laurent expansion of this new basis element reads
\begin{align}
   \masterintegralpicextra &= -\frac{1}{\epsilon^6}
   +\frac{2 \zeta_2}{\epsilon^4}
   +\frac{148 \zeta_3}{3 \epsilon^3}
   +\frac{68 \zeta_4}{\epsilon^2}
   \nonumber\\&\quad
   +\frac{\frac{3724 \zeta_5}{5}-\frac{296 \zeta_2 \zeta_3}{3}}{\epsilon}
   +\left(1491 \zeta_6-\frac{10952 \zeta_3^2}{9}\right)
   +O\left(\epsilon\right)
\end{align}

\paragraph{Dimension 3}

As mentioned above, a three-loop evaluation is hampered by the emergence of non-trivial five- and six-loop momentum-integral topologies, so we limit the calculation to two loops.

At dimension 3, the tree-level correlation function in position space reads
\begin{equation}
G_3^{(0)} = 3\, \Pi^3(x,\epsilon)\, \frac{(N^2 - 1) (N^2 - 4)}{N}
\end{equation}
At one loop, taking the ratio to the tree-level result, we obtain
\begin{align}\label{eq:twoloop3}
F_3^{(1)} &= -3\, \frac{1-3\epsilon}{1-2\epsilon}\, 
f^{(1)}
\end{align}
in terms of the uniformly transcendental function \eqref{eq:oneloopf}. This result coincides with the general formulae \eqref{eq:oneloopratio}--\eqref{eq:oneloopf}.
At two loops, the color factor is unchanged apart from an overall power of $N$, which we absorb into $\lambda$. After taking the ratio to the tree-level result and extracting the prefactor $\frac{1-4\epsilon}{1-2\epsilon}$, the final result reads
\begin{equation}
   F_3^{(2)} = 3\, \frac{1-4\epsilon}{1-2\epsilon}\, f_3^{(2)}
\end{equation}
in terms of the weight 4 uniformly transcendental function
\begin{align}\label{eq:f3}
f_3^{(2)} &=
120 \zeta _5 \epsilon +\left(432 \zeta _3^2+300 \zeta _6\right) \epsilon ^2
+\left(1296 \zeta _3 \zeta
   _4+120 \zeta _2 \zeta _5+2217 \zeta _7\right) \epsilon ^3
\nonumber\\&   
   +\left(\frac{1944 \zeta
   _{5,3}}{5}+432 \zeta _2 \zeta _3^2+9848 \zeta _5 \zeta _3+\frac{74039 \zeta
   _8}{10}\right)\epsilon ^4 
\nonumber\\&   
   +\left(5032 \zeta _3^3+24728 \zeta _6 \zeta _3+11946 \zeta _4 \zeta _5+2217 \zeta _2
   \zeta _7+32864 \zeta _9\right) \epsilon ^5
\nonumber\\&   
   +\left(\frac{1944}{5} \zeta _2 \zeta
   _{5,3}+\frac{17253 \zeta _{7,3}}{7}+22968 \zeta _4 \zeta _3^2+9848 \zeta _2 \zeta _5 \zeta
   _3+127724 \zeta _7 \zeta _3
\right.\nonumber\\&\left.   
   +\frac{262683 \zeta _5^2}{7}+\frac{187309209 \zeta
   _{10}}{1400}\right)\epsilon ^6 
\nonumber\\&   
   +\left(\frac{59616}{5} \zeta _3 \zeta _{5,3}+\frac{38448}{5} \zeta
   _{5,3,3}+5032 \zeta _2 \zeta _3^3+\frac{2856832}{15} \zeta _5 \zeta _3^2+\frac{6289739 \zeta _8
   \zeta _3}{15}
\right.\nonumber\\&\left.    
   +\frac{265053 \zeta _5 \zeta _6}{2}+\frac{3291171 \zeta _4 \zeta _7}{20}+378896 \zeta
   _2 \zeta _9-\frac{697332 \zeta _{11}}{5}\right)\epsilon ^7 
\nonumber\\& 
   +\left(\frac{47034}{5} \zeta _4 \zeta
   _{5,3}-\frac{267003}{7} \zeta _2 \zeta _{7,3}+85056 \zeta _{9,3}+13536 \zeta
   _{6,4,1,1}+\frac{128048 \zeta _3^4}{3}
\right.\nonumber\\&   
   +\frac{1526552}{3} \zeta _6 \zeta _3^2+\frac{2480482}{5}
   \zeta _4 \zeta _5 \zeta _3-7636 \zeta _2 \zeta _7 \zeta _3+\frac{6263644 \zeta _9 \zeta
   _3}{3}
\nonumber\\&\left. 
   -\frac{68949}{7} \zeta _2 \zeta _5^2+\frac{23188308 \zeta _5 \zeta
   _7}{35}+\frac{220304170443 \zeta _{12}}{110560}\right)  \epsilon ^8  
   +O\left(\epsilon ^9\right)
\end{align}
In this case, the result is directly uniformly transcendental.
The function $F_3^{(2)}$ is obtained from the following combination of master integrals:
\begin{align}
F_3^{(2)} &=
\frac{e^{-6\gamma_E\epsilon}\left(x^2\right)^{-2\epsilon}}{\Pi^{3}(x,\epsilon)}\, \frac{1}{4^{\frac52 d-7}}
\int \frac{d^d p}{\pi^{d/2}}\, e^{i p \cdot x}
\bigg[
\nonumber\\&\qquad
\frac{18 (3-4 \epsilon) (4-5 \epsilon) (3-5 \epsilon)
(83 \epsilon^2 -49\epsilon+7)}{\epsilon ^4 (1-3 \epsilon)}
\masterintegralpicN{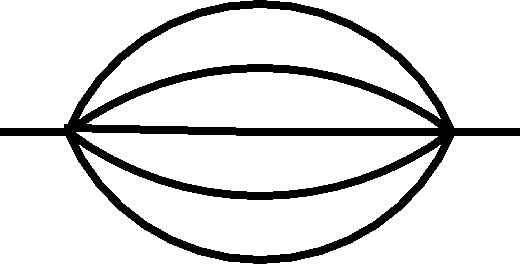}\, \frac{1}{\left(p^2\right)^2}
\nonumber\\&\qquad
+\frac{18 (1-2 \epsilon) (2-3 \epsilon )^2}{\epsilon ^2 (2-5 \epsilon)}
\masterintegralpicN{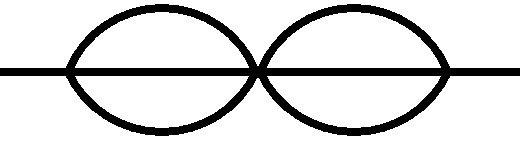}\, \frac{1}{p^2}
\nonumber\\&\qquad
+\frac{45 (1-2 \epsilon) (2-3 \epsilon) (3-5 \epsilon)}{\epsilon ^3}
\masterintegralpicN{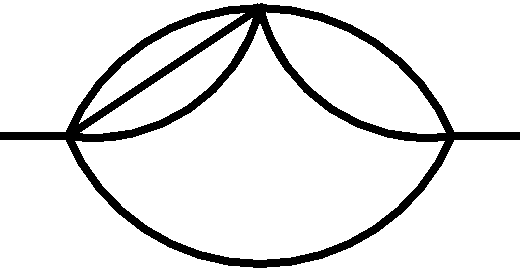}\, \frac{1}{\left(p^2\right)^2}
\nonumber\\&\qquad
+\frac{72 (1-2 \epsilon )^2 (1-4 \epsilon) (3-5 \epsilon)}{\epsilon ^3 (1-3 \epsilon)}
\masterintegralpicN{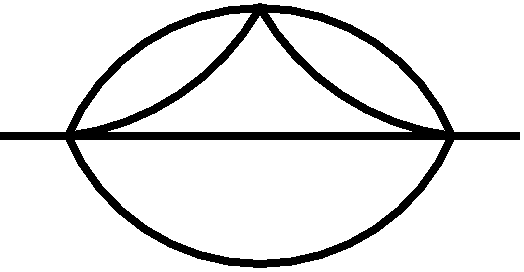}\, \frac{1}{\left(p^2\right)^2}
\nonumber\\&\qquad
+\frac{18 (1-2 \epsilon )^2}{\epsilon ^2}\masterintegralpicN{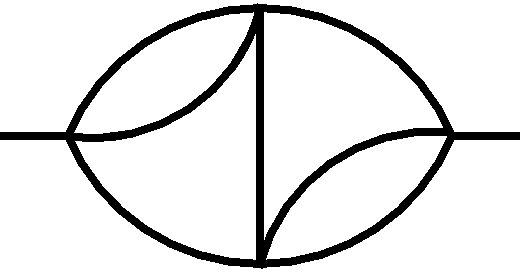}
+\frac{9 (1-2 \epsilon )^2}{\epsilon ^2}\masterintegralpicN{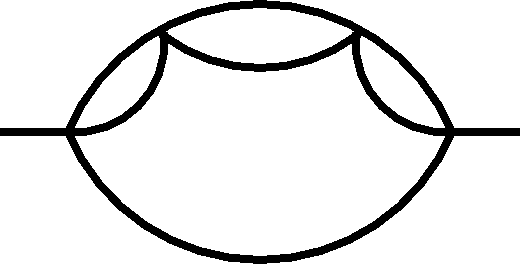}
\nonumber\\&\qquad
+\frac{27 (1-5\epsilon+6 \epsilon^2)}{2\epsilon ^2}\masterintegralpicN{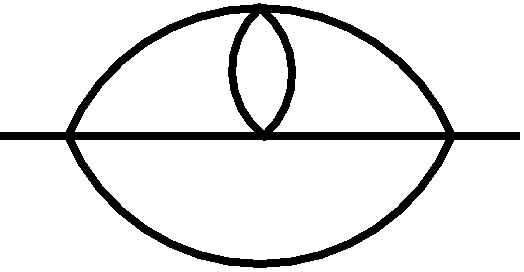}
-\frac{9 \epsilon}{2-5 \epsilon}\masterintegralpicN{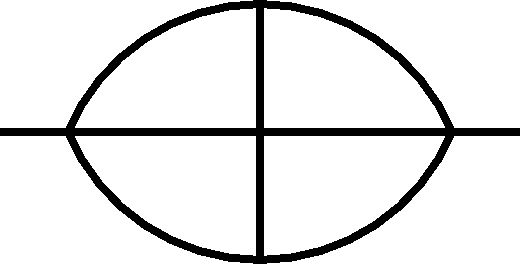}\, p^2
\bigg]
\end{align}
The first line includes the Fourier transform to position space, while the factors of $p^2$ multiplying the integrals ensure the correct dimensional scaling. In the prefactor, the free propagators $\Pi(x,\epsilon)$ come from the normalization by the tree-level result, which also removes the color factors. The factor $4^{\frac52 d-7}$ cancels an identical one from the Fourier transform; this is the price paid here for simplifying the definition \eqref{eq:twoloop3}. The factors of $(x^2)^\epsilon$ and $e^{\gamma_E\epsilon}$ are included to keep the formulae and their $\epsilon$ expansions cleaner. The rational coefficients multiplying the master integrals arise from the IBP reduction.
To make uniform transcendentality manifest, we switch to the uniformly transcendental basis proposed in \cite{Bianchi:2025sjc} and obtain
\begin{align}\label{eq:F3}
F_3^{(2)} &=  
\frac{\epsilon }{(1-5 \epsilon ) \left(1-\frac{5 \epsilon }{2}\right) (1-2 \epsilon )}\, \frac{e^{-6\gamma_E\epsilon}\left(x^2\right)^{-2\epsilon}}{\Pi^{3}(x,\epsilon)}\, \frac{1}{4^{\frac52 d-7}}\,
\int \frac{d^d p}{\pi^{d/2}}\, e^{i p \cdot x} \, \left(p^2\right)^2\, \frac{1}{16}
\bigg[
\nonumber\\&\qquad
 18\left(\masterintegralpic{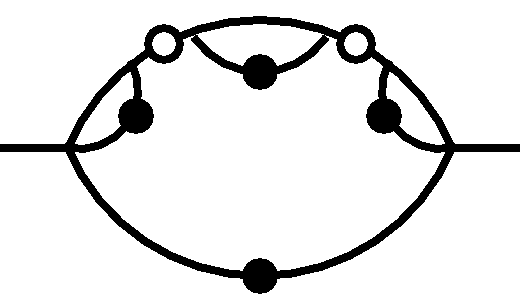}-3 \masterintegralpic{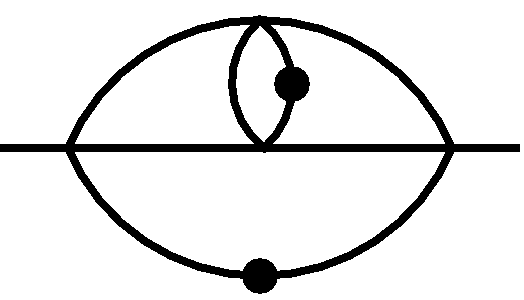}-2\masterintegralpic{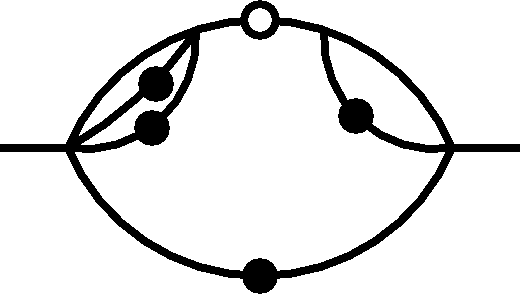}
\right)  
\nonumber\\&
\qquad + 3 \left(11 \masterintegralpic{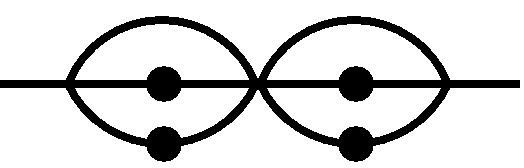}\,p^2+64 \masterintegralpic{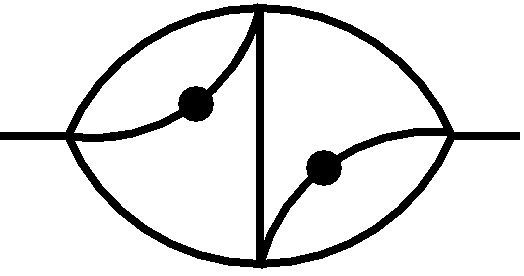}+12\masterintegralpic{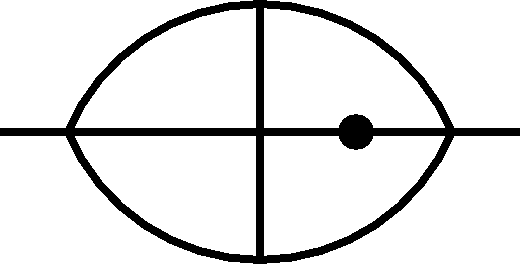}\right)
\nonumber\\&
\qquad +37 \masterintegralpic{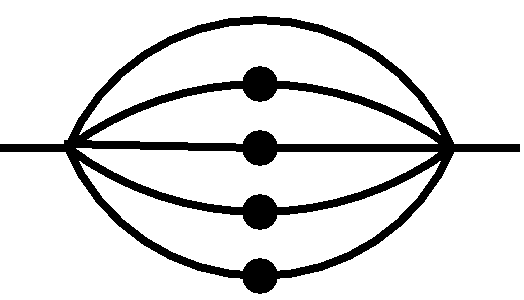}
 -52 \masterintegralpic{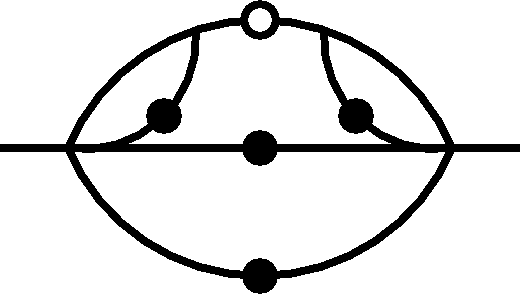}
\bigg]
\end{align}
As above, solid dots in the integral diagrams denote squared propagators, corresponding to $(k^2)^{-2}$ for momentum $k$, while empty dots denote numerator factors $k^2$.
The expression simplifies dramatically: only an overall rational function of $\epsilon$ survives, but the relative coefficients of master integrals are purely rational numbers.
This factor combines with the Fourier transform
\begin{equation}
   \frac{\epsilon }{(1-5 \epsilon ) \left(1-\frac{5 \epsilon }{2}\right) (1-2 \epsilon )}\, \frac{1}{4^{\frac52 d-7}}\, 
\int \frac{d^d p}{\pi^{d/2}}\, e^{i p \cdot x} \, \left(p^2\right)^{1-4\epsilon} = -\frac{1-4\epsilon}{1-2\epsilon}\,  \frac{ 2\epsilon\,  \Gamma (1-5 \epsilon )}{\Gamma (4 \epsilon )}\, \frac{1}{\left(x^2\right)^{3-5\epsilon}}
\end{equation}
giving rise to a uniformly transcendental in $\epsilon$ $\Gamma$-function part, while the remaining term is precisely the rational prefactor displayed above \eqref{eq:twoloop3}, which appears ubiquitously in the two-loop results presented below.

\paragraph{Ubiquitous integral combinations}
The groupings of integrals in the second, third, and fourth lines of \eqref{eq:F3} are not accidental: they identify ubiquitous combinations that recur throughout the two-loop results below. This is natural in view of the strategy described at the beginning of the section, where the calculation was organized in terms of an effective vertex based on a dimension-3-like two-point function and inserted into higher-dimensional operator correlators. As a result, the same combinations repeatedly appear with fixed relative coefficients.

It is useful to include in the definition not only the master-integral combination itself, but also the Fourier transform and the common rational factors that accompany it in the reduced correlators. We therefore define the linear functional
\begin{align}
\mathcal{P}[\mathcal{X}] &\equiv
\frac{\epsilon }{(1-5 \epsilon ) \left(1-\frac{5 \epsilon }{2}\right) (1-2 \epsilon )}\, \frac{e^{-6\gamma_E\epsilon}\left(x^2\right)^{-2\epsilon}}{\Pi^{3}(x,\epsilon)}\, \frac{1}{4^{\frac52 d-7}}
\int \frac{d^d p}{\pi^{d/2}}\, e^{i p \cdot x} \, \left(p^2\right)^2\, \mathcal{X}
\end{align}
In terms of this functional, the recurrent combinations appearing in \eqref{eq:F3} are
\begin{align}
\mathcal{A} &\equiv \mathcal{P}\left[\masterintegralpic{25}-3 \masterintegralpic{26}-2\masterintegralpic{13}\right]
\nonumber\\
\mathcal{B} &\equiv \mathcal{P}\left[11 \masterintegralpic{12}\,p^2+64 \masterintegralpic{22}+12\masterintegralpic{36}\right]
\end{align}
The same normalization will also be used below for the higher-dimensional correlators. In particular, at dimension 4 further recurring combinations appear,
\begin{align}
\mathcal{C} &\equiv \mathcal{P}\left[\masterintegralpic{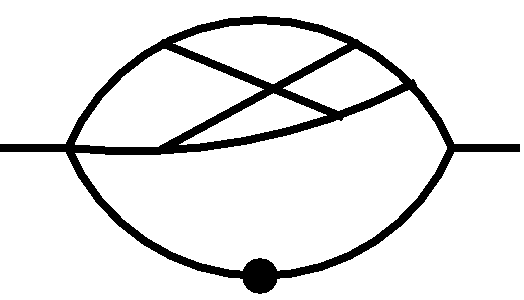}\,p^2-4\masterintegralpic{13}\right]
\nonumber\\
\mathcal{D} &\equiv \mathcal{P}\left[
   \masterintegralpic{01}-2\masterintegralpic{14}+\masterintegralpicextra
\right]
\end{align}
It is useful to name the first two integrals appearing in $\mathcal{D}$, since they will also occur independently in the following equations
\begin{align}
\mathcal{M}_{01} &\equiv \mathcal{P}\left[\masterintegralpic{01}\right]
\nonumber\\
\mathcal{M}_{04} &\equiv \mathcal{P}\left[\masterintegralpic{14}\right]
\end{align}

Then \eqref{eq:F3} can be written compactly as
\begin{equation}
   F_3^{(2)}=\frac{1}{16}\left(18\mathcal{A}+3\mathcal{B}+37\mathcal{M}_{01}-52\mathcal{M}_{04}\right).
\end{equation}
The absence of additional $\epsilon$-dependent rational coefficients in the uniformly transcendental basis of \eqref{eq:F3} then guarantees the uniform transcendental weight of $f_3^{(2)}$.

\paragraph{Dimension 4}

We now illustrate in detail the results for operators of dimension 4.
For higher-dimensional correlators, the results are stored in electronic form because of their length and to facilitate comparisons with other methods.

At dimension 4, we use the trace basis
\begin{equation}
\vec{O}_{4} = \left( \Tr(XXXX), \Tr(XX)\Tr(XX) \right)
\end{equation}

At tree level, the position-space correlation function $G_{4,i,j}^{(0)}$ reads, treating the indices $i,j$ in matrix form:
\begin{equation}
G_{4}^{(0)} = 4\, \Pi^4(x,\epsilon) \left(1-\frac{1}{N^2}\right)
\left(
\begin{array}{cc}
 N^4-6 N^2+18 & 2 N \left(2 N^2-3\right) \\
 2 N \left(2 N^2-3\right) & 2\,N^2 \left(N^2+1\right) \\
\end{array}
\right)
\end{equation}
At one loop
\begin{align}
F_4^{(1)} &= -4\, \frac{1-3\epsilon}{1-2\epsilon}\, 
f^{(1)}
\end{align}
in agreement with the general results \eqref{eq:oneloopratio}--\eqref{eq:oneloopf}.

At two loops, different trace structures begin developing different color factors with respect to their tree level analogue.
Defining by analogy with the case of dimension 3
\begin{align}
F_{4,i,j}^{(2)} &= 4\, \frac{1-4\epsilon}{1-2\epsilon}\, 
f_{4,i,j}^{(2)}
\end{align} 
The leading term in $\epsilon$ of the single-trace two-point function reads, for instance,
\begin{align}
f_{4,1,1}^{(2)} &= \frac{1}{N^4-6N^2+18}
\bigg[
60 \left(2 N^4-6 N^2+27\right) \zeta_5\, \epsilon
\\&\qquad
+6 \epsilon ^2 \bigg(
6 \left(14 N^4-82 N^2+249\right) \zeta_3^2
+25 \left(2 N^4-6 N^2+27\right) \zeta_6
\bigg)
+O\left(\epsilon ^3\right)
\bigg]\nonumber
\end{align}
At large $N$, this expression simplifies. Expanding to higher orders, we obtain
\begin{align}
f_{4,1,1}^{(2)} &= 120 \zeta_5\, \epsilon
+\left(504 \zeta_3^2+300 \zeta_6\right)\epsilon ^2
\nonumber\\&\quad
+\left(120 \zeta_2 \zeta_5+1512 \zeta_3 \zeta_4+2217 \zeta_7\right)
\epsilon ^3
\nonumber\\&\quad
+\bigg(
504 \zeta_2 \zeta_3^2+\underline{72 \zeta_3^2}+10856 \zeta_3 \zeta_5
+\frac{1944}{5}\zeta_{5,3}+\frac{75929}{10}\zeta_8
\bigg)\epsilon ^4
+O\left(\epsilon ^5\right)
\end{align}
Although the first two orders look promisingly similar to the dimension-3 case and respect uniform transcendental weight 4, deviations from uniform transcendentality start appearing at order $\epsilon^4$, as illustrated by the highlighted term in the equation.
This phenomenon occurs for all three possible correlators, $f_{4,1,1}$, $f_{4,1,2}$ and $f_{4,2,2}$, regardless of the color factors.
To inspect its origin further, we provide the expressions in terms of master integrals in the uniformly transcendental basis. For the single-trace two-point function, we find
\begin{align}
F_{4,1,1}^{(2)} &= \frac{1}{4\bigl(N^4-6N^2+18\bigr)}
\bigg[
6 N^2 (N^2+6)\left(\mathcal{A}+\frac{2 (1-3 \epsilon )^2}{3 \left(1-2\epsilon\right)\left(1-4 \epsilon\right)}\mathcal{D}\right)
\nonumber\\&\qquad
+\left(N^4-18N^2+36\right) \mathcal{B}
+18\left(2N^2-3\right) \mathcal{C}
\nonumber\\&\qquad
+\frac13 \left(37 N^4-378 N^2+900\right) \mathcal{M}_{01}
-\frac43 \left(13N^4-378 N^2+684\right) \mathcal{M}_{04}
\bigg]
\end{align}
For the mixed correlator,
\begin{align}
F_{4,1,2}^{(2)} &= \frac{1}{12\bigl(2N^2-3\bigr)}
\bigg[
90 N^2 \left(\mathcal{A}+\frac{2 (1-3 \epsilon )^2}{3 \left(1-2\epsilon\right)\left(1-4 \epsilon\right)}\mathcal{D}\right)
\nonumber\\&\qquad
-3\left(N^2+6\right) \mathcal{B}
+27\left(N^2+1\right) \mathcal{C}
\nonumber\\&\qquad
+5 \left(7 N^2-30\right)\mathcal{M}_{01}
+4 \left(49 N^2+114\right) \mathcal{M}_{04}
\bigg]
\end{align}
For the double-trace correlator,
\begin{align}
F_{4,2,2}^{(2)} &= \frac{1}{12\bigl(N^2+1\bigr)}
\bigg[
36 \left(N^2+6\right) \left(\mathcal{A}+\frac{2 (1-3 \epsilon )^2}{3 \left(1-2\epsilon\right)\left(1-4 \epsilon\right)}\mathcal{D}\right)
\nonumber\\&\qquad
-30 \mathcal{B}
+9\left(N^2+11\right) \mathcal{C}
\nonumber\\&\qquad
+2 \left(12 N^2-53\right)\mathcal{M}_{01}
+8 \left(6 N^2+131\right) \mathcal{M}_{04}
\bigg]
\end{align}
As in the dimension-3 case, these expressions exhibit recurrent combinations of integrals. At dimension 4, this pattern is enriched by an additional combination generated by the new diagrams and non-planar topologies that first contribute at this order.
From these results we see that uniform transcendentality is broken by the $\epsilon$-dependent prefactors multiplying the basis integrals. 
We do not have a deep explanation of this phenomenon. Since there is also no clear reason to expect uniform transcendentality for these quantities, it is perhaps surprising that it occurs for dimensions 2 and 3, and up to order $\epsilon^3$ for higher-dimensional contributions.

Interestingly, thanks to the particular grouping of integrals chosen here and to the seemingly baroque, but ultimately useful, definition of the five-loop integral in \eqref{eq:extra}, the breaking of uniform transcendentality is confined to specific lines of these formulae. This means that it can be removed easily by defining suitable combinations of correlators.
Specifically, at dimension 4, the linear combination displaying uniform transcendentality is
\begin{equation}
   U_4 \equiv F_{4,1,1} - \frac{N^4+N^2}{2 N^4-12 N^2+36}\, F_{4,2,2} = F_{4,1,1} - \frac12\, F_{4,2,2} + O\left(N^{-2}\right)
\end{equation}
However, this rather complicated combination does not seem to possess any deeper intrinsic physical significance.

This dimension-four example illustrates a phenomenon that turns out to be ubiquitous in the two-point functions of higher-dimensional operators, as discussed in the following section.

We conclude this section by mentioning the checks performed on these results: gauge invariance, verified by working in a generic $R_\xi$ gauge, and the fact that the two-point functions do not renormalize and are of order $\epsilon$. 

Finally, the first order in the $\epsilon$ expansion agrees with a localization prediction of \cite{AleCongkao}. This is a highly non-trivial and satisfying result.
No agreement beyond leading order can be expected with localization, since the latter calculation is performed on a sphere and then mapped to flat space by means of a conformal transformation, which applies strictly in four dimensions. By contrast, our calculation is performed in $d=4-2\epsilon$, where conformal invariance is explicitly broken.

\section{Two-loop extrapolations}

At two loops, extracting results for general dimension $n$ is more difficult, because the two-point functions do not depend on a single universal function. Instead, they depend on the specific operators in the trace basis and involve non-trivial color factors.
Still, we can extract conjectures in certain limits and subcases.

To this end, we extended the calculation outlined above for dimensions 3 and 4 to higher-dimensional operators, carrying out the full two-loop computation of all possible two-point functions up to dimension 10.
The resulting expressions are lengthy and not particularly illuminating, so we do not display them explicitly here.
For future reference, they are included with this article in electronic \texttt{Mathematica} format.
The file \texttt{data.m} contains all the relevant data, and the notebook \texttt{example.nb} illustrates how to retrieve them. In particular, the functions $f_{n,i,j}$ and $F_{n,i,j}$ are implemented at one and two loops, for $3\leq n\leq10$, both in terms of uniformly transcendental master integrals and expanded in $\epsilon$ up to transcendental order 12.

The main purpose of this calculation was to provide a database for comparisons with localization-based results, which agree up to the leading term in $\epsilon$ for all cases we checked.

Moreover, a sufficiently large data set sometimes makes it possible to extrapolate to formulae valid at generic dimension, as we now describe.

For instance, based on the perturbative results in the ancillary files, we define
\begin{equation}
F_{n,i,j}^{(2)} = n\, \frac{1-4\epsilon}{1-2\epsilon}\, 
f_{n,i,j}^{(2)}
\end{equation} 
and conjecture that the leading terms in $\epsilon$ and $N^2$ for the corresponding single-trace function are, for $n\geq 3$,
\begin{align}
f_{n,1,1}^{(2)} &= 120\,\zeta_5\,\epsilon
+12\left(6(n+3)\zeta_3^2+25\zeta_6\right)\epsilon^2
\\&\quad
+3\left(72(n+3)\zeta_3\zeta_4+40\zeta_2\zeta_5+739\zeta_7\right)\epsilon^3+O\left(\epsilon^4\right)
\nonumber\\&\quad
+\left(n^3+11n-36\right)\bigg[
5\zeta_5\,\epsilon
+\frac{1}{2}\left(2\zeta_3^2+25\zeta_6\right)\epsilon^2
\nonumber\\&\qquad
+\left(3\zeta_3\zeta_4+5\zeta_2\zeta_5+60\zeta_7\right)\epsilon^3+O\left(\epsilon^4\right)
\bigg]N^{-2}
+O\left(N^{-4}\right)\nonumber
\end{align}
For single-trace operators, the data up to $n=10$ suggest the following master-integral expression for the full two-loop correction:
\begin{align}\label{eq:Fn11}
F_{n,1,1}^{(2)} &= \frac{n}{16}\bigg[
6\mathcal{A}+\mathcal{B}
+4(n-3)\frac{(1-3 \epsilon )^2}{\left(1-2\epsilon\right)\left(1-4 \epsilon\right)}\mathcal{D}
+\frac13 \left(37\mathcal{M}_{01}+52\mathcal{M}_{04}\right)
\nonumber\\&\quad
+\frac{n^3+11n-36}{6 N^2}\bigg(
6\mathcal{A}-\mathcal{B}+3\mathcal{C}
+6\frac{2(1-3 \epsilon )^2}{3\left(1-2\epsilon\right)\left(1-4 \epsilon\right)}\mathcal{D}
\nonumber\\&\qquad\qquad
-\frac13 \left(13\mathcal{M}_{01}-100\mathcal{M}_{04}\right)
\bigg)
\bigg]
+O\left(N^{-4}\right)
\end{align}
These formulae do not interpolate smoothly to $n=2$. This is natural from the perturbative point of view: two-point functions of such short operators receive a different set of two-loop contributions, and some diagrams that would otherwise be non-planar become leading in the color expansion because of the special color algebra of a trace of two fields. This is perhaps loosely reminiscent of wrapping effects in integrability, in the sense that the shortest operators require separate treatment rather than following the generic-length pattern. A clearer understanding of this failure emerges in the planar limit, as discussed in section \ref{sec:full-planar}. 
After formulating this prediction, we pushed the calculation of single-trace operators to $n=11$ with a standalone calculation of $F_{11,1,1}^{(2)}$, which agrees with this conjecture.

Higher non-planar corrections are more challenging to fit with a general ansatz, although this should be possible by pushing the calculation to higher $n$.

One special case where the full expression at finite $N$ can be extrapolated for any $n\geq 4$ is that of products $O_{2p, -1}=O_{2}^p$ (-1 stands here as the cyclic index like in \texttt{Mathematica}).
Then the first few orders in the $\epsilon$ expansion of $f_{2p,-1,-1}^{(2)}$ read
\begin{align}\label{eq:Fnm1m1even}
f_{2p,-1,-1}^{(2)} &=
\frac{150\,\zeta_5\left(N^2+2p-1\right)}{N^2+1}\,\epsilon
\nonumber\\&\quad
+\frac{3}{N^2+1}\left[
\zeta_3^2\left(N^2(48p+74)+68p+54\right)
+125\zeta_6\left(N^2+2p-1\right)
\right]\epsilon^2
\nonumber\\&\quad
+\frac{3}{N^2+1}\bigg[
50\zeta_2\zeta_5\left(N^2+2p-1\right)
+6\zeta_3\zeta_4\left(N^2(24p+37)+34p+27\right)
\nonumber\\&\qquad
+\zeta_7\left(859N^2+1200p-341\right)
\bigg]\epsilon^3
+O\left(\epsilon^4\right)
\end{align}
and more orders can in principle be obtained by expanding the following general expression in terms of uniformly transcendental master integrals:
\begin{align}
F_{2p,-1,-1}^{(2)} &= \frac{n}{N^2+1}\bigg[
6 \left(5 n+2 N^2-8\right)\mathcal{A}
-5(n-2)\mathcal{B}
+3 \left(5 n+N^2-9\right)\mathcal{C}
\nonumber\\&\qquad
+4(n-2)(N^2+6)\frac{(1-3 \epsilon )^2}{\left(1-2\epsilon\right)\left(1-4 \epsilon\right)}\mathcal{D}
\nonumber\\&\qquad
+\frac{-65 n+24 N^2+154}{3}\mathcal{M}_{01}
+\frac43 \left(125 n+12 N^2-238\right)\mathcal{M}_{04}
\bigg]
\end{align}
For odd $(2p+1)$-dimension operators, the operators with the highest number of multiple traces possess the form $O_{3}\, O_{2}^{p-1}$. In this case too we can extrapolate general results, for $n\geq 5$
\begin{align}\label{eq:Fnm1m1odd}
n\, f_{2p+1,-1,-1}^{(2)} &=
\bigg[
\frac{2520 (p-1)^2 \zeta_5}{N^2+5}
-\frac{1920 (p-2)(p-1) \zeta_5}{N^2+7}
+60(5p+1)\zeta_5
\bigg]\epsilon
\nonumber\\&\quad
+\bigg[
\frac{252 (p-1)^2\left(2\zeta_3^2+25\zeta_6\right)}{N^2+5}
-\frac{192 (p-2)(p-1)\left(2\zeta_3^2+25\zeta_6\right)}{N^2+7}
\nonumber\\&\qquad
+12\left(24p^2+61p+23\right)\zeta_3^2
+150(5p+1)\zeta_6
\bigg]\epsilon^2
\nonumber\\&\quad
+\bigg[
\frac{504 (p-1)^2\left(5\zeta_2\zeta_5+3\zeta_3\zeta_4+60\zeta_7\right)}{N^2+5}
\nonumber\\&\qquad
-\frac{384 (p-2)(p-1)\left(5\zeta_2\zeta_5+3\zeta_3\zeta_4+60\zeta_7\right)}{N^2+7}
+60(5p+1)\zeta_2\zeta_5
\nonumber\\&\qquad
+36\left(24p^2+61p+23\right)\zeta_3\zeta_4
+3(1718p+499)\zeta_7
\bigg]\epsilon^3
+O\left(\epsilon^4\right)
\end{align}
from the master integrals expression
\begin{align}
F_{2p+1,-1,-1}^{(2)} &= \frac{n-3}{16}\bigg[
6 \left(-\frac{16 (n-5)}{N^2+7}+\frac{21(n-3)}{N^2+5}+\frac{n}{n-3}+1\right)\mathcal{A}
\nonumber\\&\quad
+\left(-\frac{16 (n-5)}{N^2+7}+\frac{21 (n-3)}{N^2+5}-\frac{3}{n-3}\right)\mathcal{B}
\nonumber\\&\quad
+3 \left(-\frac{16 (n-5)}{N^2+7}+\frac{21 (n-3)}{N^2+5}+1\right)\mathcal{C}
\nonumber\\&\quad
+4 \left(-\frac{16 (n-5)}{N^2+7}+\frac{21 (n-3)}{N^2+5}+n+1\right)
\frac{(1-3 \epsilon )^2}{\left(1-2\epsilon\right)\left(1-4 \epsilon\right)}\mathcal{D}
\nonumber\\&\quad
+\frac{1}{3}\left(\frac{208 (n-5)}{N^2+7}-\frac{273(n-3)}{N^2+5}+\frac{111}{n-3}+24\right)
\mathcal{M}_{01}
\nonumber\\&\quad
+4 \left(-\frac{400 (n-5)}{3 \left(N^2+7\right)}+\frac{175 (n-3)}{N^2+5}-\frac{13}{n-3}+4\right)
\mathcal{M}_{04}
\bigg]
\end{align}
In the planar limit, we present further extrapolations in section \ref{sec:full-planar}.

\section{Uniformly transcendental subtractions}
From these results, we can also explicitly confirm the expectation that, in the planar limit, a suitable combination of the single-trace and maximally factorized operators gives rise to a uniformly transcendental combination for general $n$.
Requiring the cancellation of the $\epsilon$ dependence in a linear combination of \eqref{eq:Fn11} and either \eqref{eq:Fnm1m1even} or \eqref{eq:Fnm1m1odd} fixes the relative coefficient between the single-trace and maximal multi-trace operators to be
\begin{equation}
   a_n = \frac{1+(-1)^n}{2}\,\frac{3-n}{n-2} - \frac{1-(-1)^n}{2}\,\frac{n}{n+1} \qquad\qquad n\geq4
\end{equation}
which for even $n$ gives the simple sequence
\begin{equation}
   a_{2n}= -\left(\frac12,\, \frac34,\, \frac56,\, \frac78,\, \dots  \right)
\end{equation}
and similarly for odd $n$, starting at $\frac56$.
Since the relative coefficients of the two-point functions are negative, these combinations cannot be interpreted as consistent two-point-function norms of linear combinations of operators, such as those emerging from an orthogonalization procedure, for which the relative coefficients between correlators would be non-negative.
Rather, these uniformly transcendental combinations are closer in spirit to cumulant-like subtractions of normalized correlator data. They are not operator-level projections; instead, they partially remove specific multi-trace contributions with special coefficients.

Starting at dimension 6, several multi-trace operators appear, allowing for more combinations that preserve uniform transcendentality.
These can be identified as follows. At the level of the master-integral expressions, a simple criterion for uniform transcendentality is the cancellation of the five-loop integral contribution. The coefficients enforcing this cancellation automatically leave the coefficients of all remaining integrals independent of $\epsilon$, and therefore the result is uniformly transcendental.

Using this criterion, we calculated the coefficients of the offending integral in all two-point functions of an operator with itself, which are the only non-vanishing ones in the planar limit. These coefficients define a rational vector $\vec{v}_n$ in a space whose dimension is the number of integer partitions of $n$.
We then constructed bases for the spaces orthogonal to these vectors. These bases encode the relative coefficients of the various two-point functions of dimension-$n$ operators that enforce uniform transcendentality.
The notable result is that, up to an overall constant, the resulting combinations of master integrals are all equivalent at fixed $n$.
Furthermore, they can be predicted on general grounds.
This is because the vectors $\vec{v}_n$ themselves obey a simple combinatorial pattern associated with the partition $\Lambda$ corresponding to the multi-trace operator. Namely, fixing their normalization so that the single-trace operator corresponds to a unit entry, we find
\begin{equation}\label{eq:v}
   \vec{v}_n(\Lambda) = 1+ \frac{m_2(\Lambda)}{n(n-3)}
\end{equation}
where the function $m_2$ is defined as
\begin{equation}\label{eq:number-of-twos}
   m_2(\Lambda) \equiv \sum_{r=1}^{\ell(\Lambda)} \delta_{\Lambda_r,2}
   \,,\qquad
   \Lambda=(\Lambda_1,\ldots,\Lambda_{\ell(\Lambda)})\vdash n\,,\quad \Lambda_r\geq 2\,.
\end{equation}
Thus $m_2(\Lambda)$ simply counts the number of parts equal to two in the partition, or equivalently the number of $\Tr(X^2)$ factors in the corresponding multi-trace operator. These factors are the components of the stress-tensor multiplet in the field-theory description. Their distinguished role is consistent with the tower structure that appears in localization computations of protected correlators: after resolving sphere-mixing by Gram--Schmidt orthogonalization, operators organize into towers generated by repeated insertions of the dimension-2 operator, and different towers decouple \cite{Gerchkovitz:2016gxx,Brown:2023zbr,AleCongkao}.
In holographic terms, factors of $\Tr(X^2)$ correspond to insertions of the $k=2$ Kaluza--Klein mode belonging to the stress-tensor multiplet, which occupy a distinguished role among the protected supergravity excitations of the compactification.
The above formula makes the orthogonal hyperplane predictable. It turns out that, for a given dimension $n$, the uniformly transcendental combination evaluates uniquely, after choosing a convenient normalization, to
\begin{align}\label{eq:Unmaster}
U_n(\epsilon) &= 
2 \left(n-4\right)\mathcal{A}
-\frac13(n-2)\mathcal{B}
+\left(n-3\right)\mathcal{C}
+\frac19(2-13n)\mathcal{M}_{01}
+\frac49 \left(25 n-62\right)\mathcal{M}_{04}
\end{align}
whose first few expansion coefficients are
\begin{align}\label{eq:Unexpansion}
U_n(\epsilon) &=-20(n-7)\zeta_5\,\epsilon^{-1}
+\left[4 (11 n+39) \zeta_3^2-50 (n-7) \zeta_6\right]
\nonumber\\&\quad
+\bigl[
40 (n-7) \zeta_2 \zeta_5+12 (11 n+39) \zeta_3 \zeta_4+(2198-240 n)
   \zeta_7
\bigr]\epsilon
\nonumber\\&\quad
+\biggl[
-8 (11 n+39)\zeta_2 \zeta_3^2
+\frac{8}{3} (991 n-2371) \zeta_3 \zeta_5
+\frac{2592 n}{5}\zeta_{5,3}
\nonumber\\&\qquad
+\frac{1296}{5}(2n-5) \zeta_{5,3}
+\frac{265285-49069n}{30}\zeta_8
\biggr]\epsilon^2
\nonumber\\&\quad
+\biggl[
2(120 n-1099)\zeta_2 \zeta_7
+2(269-11 n)\zeta_4 \zeta_5
+\frac{2}{9}(67263-5989 n)\zeta_9
\nonumber\\&\qquad
-\frac{8}{3}(927 n+2657)\zeta_3^3
+\frac{2}{3}(8887 n-27337)\zeta_3 \zeta_6
\biggr]\epsilon^3
+O\left(\epsilon^4\right)
\end{align}
This expression is evaluated directly in momentum space, without the Fourier transform or rational correction factors that enter the perturbative ratios $F_n$.
In \eqref{eq:v}, we observe the special role of parts of size two in the partitions, corresponding to factors of $\Tr(X^2)$ in the operators. This appearance is intuitively reasonable from the perturbative perspective, since the sources of non-uniform transcendentality seem to be connected to diagrams that factorize into lower-loop two-point functions. Nevertheless, the precise coefficient needed to remove the transcendentality-breaking terms is not completely transparent.
Moreover, the presence of special factorized diagrams with $O_2$-like two-point functions might be a two-loop artifact, while at higher loops and higher operator dimensions more factorization structures become available. It would be interesting to check whether, and how, uniformly transcendental quantities can still be defined systematically. Such calculations are currently beyond our capabilities.

The analysis above was restricted to the planar limit for simplicity. In the non-planar case, more mixing occurs, but it is still possible to engineer subtractions with coefficients that are rational functions of $N$ so as to construct uniformly transcendental combinations. These nevertheless turn out to be equivalent, up to an overall factor, to those described above in the planar case.

\section{Full planar extrapolation at two loops}\label{sec:full-planar}

Inspired by the results above and the crucial role played by parts of size two in the partitions defining the trace structure, we formulate a full conjecture for the two-loop planar corrections to two-point functions; in the planar limit, only correlators with the same trace structure survive. In terms of master integrals, it takes the following form:
\begin{align}\label{eq:fullplanar}
F_{n,\Lambda,\Lambda}^{(2)} &= n\bigg[
\frac{3}{8}\left(1+\frac{2m_2}{n}\right)\mathcal{A}
+\frac{1}{16}\left(1-\frac{2m_2}{n}\right)\mathcal{B}
+\frac{3m_2}{8n}\mathcal{C}
\nonumber\\&\qquad
+\frac{(1-3 \epsilon)^2}{(1-4 \epsilon)(1-2 \epsilon)}
\left(\frac{m_2}{2 n}+\frac{n-3}{4}\right)\mathcal{D}
\nonumber\\&\qquad
+\frac{1}{24}\left(\frac{37}{2}-\frac{13m_2}{n}\right)
\mathcal{M}_{01}
+\frac{1}{6}\left(\frac{25m_2}{n}-\frac{13}{2}\right)
\mathcal{M}_{04}
\bigg]
\end{align}
Here $\Lambda$ denotes the partition specifying the trace structure of the operator, and $m_2 \equiv m_2(\Lambda)$ is the number of parts of $\Lambda$ equal to 2, as defined in \eqref{eq:number-of-twos}.
The appearance of this grading in the leading $\epsilon$ term is consistent with, and perhaps anticipated by, the localization tower structure mentioned above \cite{Brown:2023zbr}.\footnote{I thank Congkao Wen and Alessandro Georgoudis for correspondence pointing out this connection.} What is less automatic is the stronger statement visible in \eqref{eq:fullplanar}: the same $m_2$-grading controls the full planar two-loop reduced correlator, including the coefficients of the individual master-integral combinations and all higher terms in the $\epsilon$ expansion. This appears to be a special feature of the planar limit.
After Fourier transforming to position space and using the reduced normalization introduced above, the first terms of the corresponding $\epsilon$ expansion are
\begin{align}
f_{n,\Lambda,\Lambda}^{(2)} &=
60\left(2+\frac{m_2}{n}\right)\zeta_5\,\epsilon
+\bigg[
12\left(\frac{m_2}{n}+6(n+3)\right)\zeta_3^2
+150\left(2+\frac{m_2}{n}\right)\zeta_6
\bigg]\epsilon^2
\nonumber\\&
+\bigg[
60\left(2+\frac{m_2}{n}\right)\zeta_2\zeta_5
+36\left(\frac{m_2}{n}+6(n+3)\right)\zeta_3\zeta_4
+\left(2217+\frac{720m_2}{n}\right)\zeta_7
\bigg]\epsilon^3
\nonumber\\&
+\bigg[
72\left(n-3+\frac{2m_2}{n}\right)\zeta_3^2 + 12\left(\frac{m_2}{n}+6n+18\right)\zeta_2\zeta_3^2
\nonumber\\&\qquad
+8\left(126n+853-\frac{274m_2}{n}\right)\zeta_3\zeta_5
+\frac15\left(1944-\frac{7776m_2}{n}\right)\zeta_5\zeta_3
\nonumber\\&\qquad
+\left(189n+\frac{68369}{10}+\frac{60349m_2}{10n}\right)\zeta_8
\bigg]\epsilon^4
+O\left(\epsilon^5\right)
\end{align}

\section{Conclusions}

In this paper we have studied two-point functions of protected scalar operators in $\mathcal{N}=4$ SYM, focusing on their transcendentality properties in dimensional regularization. We worked in dimensional reduction and organized the two-loop calculation in terms of effective vertices and propagator-type master integrals. This allowed us to compute the relevant correlators for all trace structures up to dimension 10, to extract compact formulae in several infinite families, and to formulate a full planar extrapolation for diagonal two-loop correlators at arbitrary dimension and trace structure.

At one loop, we found a universal answer: after factoring out the tree-level correlator, the dependence on the operator dimension is entirely captured by an overall factor. The remaining function is uniformly transcendental, in agreement with localization-based expectations for the first terms in the $\epsilon$ expansion. At two loops, the dimension-3 single-trace correlator continues to display uniform transcendentality after a natural normalization. This provides a non-trivial extension of the pattern previously observed for dimension-2 operators.

Starting at dimension 4, individual two-point functions no longer remain uniformly transcendental. The breaking, however, is highly structured. In the uniformly transcendental basis of master integrals, the offending terms are confined to specific combinations, closely tied to the five-loop integral introduced in section \ref{sec:two-loop}. This makes it possible to remove them by taking suitable linear combinations of correlators. At dimension 4, this already produces a simple combination of single- and double-trace correlators, and the same mechanism extends to higher dimensions.

Using the data up to dimension 10, we identified a simple planar pattern controlling these uniformly transcendental subtractions. The relevant coefficients are encoded in the vector \eqref{eq:v}, whose dependence on the trace structure is governed by the number of factors of $\Tr(X^2)$ in the corresponding partition. This suggests that the sources of non-uniform transcendentality are associated with factorized lower-dimensional two-point-function substructures. One of the main outcomes of the paper is that the same counting function controls more than the subtraction coefficients: in the planar limit, the diagonal two-loop correlator for a general partition $\Lambda$ admits the compact master-integral formula of section \ref{sec:full-planar}, and its perturbative expansion depends on the trace structure only through $m_2(\Lambda)$. The non-planar case involves more mixing, but the resulting uniformly transcendental combinations appear to be equivalent, up to an overall factor, to the planar ones.

Several questions remain open. It would be interesting to find a more conceptual derivation of the combinatorial pattern behind \eqref{eq:v}, and of the stronger $m_2(\Lambda)$ dependence of the full planar correlator. It would also be useful to determine whether analogous uniformly transcendental combinations and privileged $O_2$ structures persist at higher loops, where new factorization channels and more complicated operator mixing are expected to appear. 
Localization suggests that such an $m_2$ dependence should persist at least in the leading term of the $\epsilon$ expansion, since it reflects a general feature of the organization of protected operators. From this perspective, the special role of $O_2$ is natural: it belongs to the stress-tensor multiplet and, in localization, is directly related to derivatives of the partition function with respect to the coupling. A derivation of this structure could also clarify the underlying holographic interpretation.  Finally, the data presented here provide further perturbative input for comparison with localization and may help clarify how four-dimensional protected correlators are encoded in dimensionally regularized perturbation theory. Similar questions could be explored in other models, such as ABJM theory, where localization is also available but protected operators receive finite quantum corrections, and where related two- and three-point functions have already been studied in perturbation theory \cite{Young:2014lka,Young:2014sia,Bianchi:2020cfn,Bianchi:2024nah}.

\acknowledgments

I thank Alessandro Georgoudis, Joe Minahan, Anton Nedelin and Congkao Wen for sharing their localization-based results, which provided substantial motivation for this work. I am especially grateful to Alessandro Georgoudis and Congkao Wen for correspondence and discussions.
This work was supported by Fondo Nacional de Desarrollo Cient\'ifico y Tecnol\'ogico, through Fondecyt Regular 1220240, Fondecyt Exploraci\'on 13220060 and Fondecyt Exploraci\'on 13250014.

\vfill
\newpage

\appendix

\section{Master integral expansions}\label{app:expansions}

For Feynman integrals, we use the integration measure $\int \frac{e^{\gamma_E \epsilon}\, d^{4-2\epsilon} k}{\pi^{2-\epsilon}}$.
For the one-loop corrections, we use the following integrals:
\begin{alignat}{2}
&\loopintegralpic{M2L1} &&= \frac{e^{2 \gamma_E \epsilon}\, \Gamma (3-d) \Gamma \left(\frac{d}{2}-1\right)^3}{\Gamma \left(\frac{3d}{2}-3\right)}
\\
&\loopintegralpic{M2L2} &&= \frac{e^{2 \gamma_E \epsilon}\, \Gamma \left(2-\frac{d}{2}\right)^2 \Gamma \left(\frac{d}{2}-1\right)^4}{\Gamma (d-2)^2}
\end{alignat}
We use the Fourier transform
\begin{equation}
\int \frac{d^d p}{\pi^{d/2}}\, e^{i p\cdot x}\,(p^2)^{\alpha}
=
4^{\frac d2+\alpha}
\frac{\Gamma\!\left(\frac d2+\alpha\right)}
{\Gamma(-\alpha)}
\frac{1}{(x^2)^{\frac d2+\alpha}}
\label{eq:FTconvention}
\end{equation}
The four-loop uniformly transcendental master integrals use the following graphical convention: squared propagators are represented by solid dots and inverse propagators by empty dots. For instance,
\begin{equation}
   \masterintegralpic{14} = \int \frac{e^{4 \gamma_E \epsilon}\, d^d k_1\, d^d k_2\, d^d k_3\, d^d k_4\, k_2^2}{(\pi^{d/2})^4\, k_1^2 \left((k_1-k_2)^2\right)^2 k_3^2 \left((k_2-k_3)^2\right)^2 \left((k_2-k_4)^2\right)^2 \left((k_4-p)^2\right)^2}
\end{equation}
We report the Laurent expansions for the four-loop masters needed for the two-loop calculations of section \ref{sec:two-loop}.
The integrals are evaluated at $p^2=1$. Only a few orders are displayed. The full expressions up to transcendental order 12, taken from \cite{Bianchi:2025sjc} and based on \cite{Baikov:2010hf,Lee:2011jt}, are stored in the ancillary file \texttt{data.m}, both in the following form and after Fourier transformation to position space and multiplication by $\frac{1-4\epsilon}{1-2\epsilon}$ to preserve uniform transcendentality and contribute directly to the reduced correlators $f_{n,i,j}$. 
\begin{alignat}{2}
&\masterintegralpic{01} &&= -\frac{1}{\epsilon ^6}
   +\frac{2 \zeta_2}{\epsilon ^4}
   +\frac{184 \zeta_3}{3 \epsilon ^3}
   +\frac{86 \zeta_4}{\epsilon^2}
   \nonumber\\& &&\quad
   +\frac{\frac{4144 \zeta_5}{5}-\frac{368 \zeta_2 \zeta_3}{3}}{\epsilon }
   +\left(1608 \zeta_6-\frac{16928 \zeta_3^2}{9}\right)
   +O\left(\epsilon\right)
   \\[0.5em]
&\masterintegralpic{12} &&= -\frac{1}{\epsilon ^6}
   +\frac{2 \zeta_2}{\epsilon ^4}
   +\frac{64 \zeta_3}{3 \epsilon ^3}
   +\frac{26 \zeta_4}{\epsilon^2}
   \nonumber\\& &&\quad
   +\frac{\frac{544 \zeta_5}{5}-\frac{128 \zeta_2 \zeta_3}{3}}{\epsilon }
   +\left(118 \zeta_6-\frac{2048 \zeta_3^2}{9}\right)
   +O\left(\epsilon\right)
   \\[0.5em]
&\masterintegralpic{13} &&= -\frac{1}{\epsilon ^6}
   +\frac{2 \zeta_2}{\epsilon ^4}
   +\frac{136 \zeta_3}{3 \epsilon ^3}
   +\frac{62 \zeta_4}{\epsilon^2}
   \nonumber\\& &&\quad
   +\frac{\frac{3184 \zeta_5}{5}-\frac{272 \zeta_2 \zeta_3}{3}}{\epsilon }
   +\left(1252 \zeta_6-\frac{9248 \zeta_3^2}{9}\right)
   +O\left(\epsilon\right)
   \\[0.5em]
&\masterintegralpic{14} &&= -\frac{1}{\epsilon ^6}
   +\frac{2 \zeta_2}{\epsilon ^4}
   +\frac{166 \zeta_3}{3 \epsilon ^3}
   +\frac{77 \zeta_4}{\epsilon^2}
   \nonumber\\& &&\quad
   +\frac{\frac{3934 \zeta_5}{5}-\frac{332 \zeta_2 \zeta_3}{3}}{\epsilon }
   +\left(\frac{3099 \zeta_6}{2}-\frac{13778 \zeta_3^2}{9}\right)
   +O\left(\epsilon\right)
   \\[0.5em]
&\masterintegralpic{22} &&= \frac{6 \zeta_3}{\epsilon ^3}
   +\frac{9 \zeta_4}{\epsilon^2}
   +\frac{127 \zeta_5-12 \zeta_2 \zeta_3}{\epsilon}
   +\left(271 \zeta_6-229 \zeta_3^2\right)
   +O\left(\epsilon\right)
   \\[0.5em]
&\masterintegralpic{25} &&= -\frac{1}{\epsilon ^6}
   +\frac{2 \zeta_2}{\epsilon ^4}
   +\frac{118 \zeta_3}{3 \epsilon ^3}
   +\frac{53 \zeta_4}{\epsilon^2}
   \nonumber\\& &&\quad
   +\frac{\frac{2974 \zeta_5}{5}-\frac{236 \zeta_2 \zeta_3}{3}}{\epsilon }
   +\left(\frac{2387 \zeta_6}{2}-\frac{6962 \zeta_3^2}{9}\right)
   +O\left(\epsilon\right)
   \\[0.5em]
&\masterintegralpic{26} &&= \frac{6 \zeta_3}{\epsilon ^3}
   +\frac{9 \zeta_4}{\epsilon^2}
   +\frac{102 \zeta_5-12 \zeta_2 \zeta_3}{\epsilon}
   +\left(\frac{417 \zeta_6}{2}-314 \zeta_3^2\right)
   +O\left(\epsilon\right)
   \\[0.5em]
&\masterintegralpic{36} &&= -\frac{20 \zeta_5}{\epsilon }
   +\left(-32 \zeta_3^2-50 \zeta_6\right)
   +O\left(\epsilon\right)
   \\[0.5em]
&\masterintegralpic{43} &&= -\frac{20 \zeta_5}{\epsilon }
   +\left(-68 \zeta_3^2-50 \zeta_6\right)
   +O\left(\epsilon\right)
   \\[0.5em]
&\masterintegralpicextra &&= -\frac{1}{\epsilon ^6}
   +\frac{2 \zeta_2}{\epsilon ^4}
   +\frac{148 \zeta_3}{3 \epsilon ^3}
   +\frac{68 \zeta_4}{\epsilon^2}
   \nonumber\\& &&\quad
   +\frac{\frac{3724 \zeta_5}{5}-\frac{296 \zeta_2 \zeta_3}{3}}{\epsilon }
   +\left(1491 \zeta_6-\frac{10952 \zeta_3^2}{9}\right)
   +O\left(\epsilon\right)
\end{alignat}

\bibliographystyle{JHEP}
\bibliography{biblio3}

\end{document}